\documentclass[12pt]{article}

\usepackage[margin=1in]{geometry}

\usepackage{amsmath, graphicx, bm, flexisym, lscape, multirow}
\usepackage{authblk}
\usepackage[round]{natbib}
\usepackage{xr}
\usepackage{setspace}
\usepackage{changebar}

\newcommand{\nc}{\newcommand}

\nc{\bbeta}{\mbox{\boldmath{$\beta$}}} 
\nc{\bxi}{\mbox{\boldmath{$\xi$}}} 
\nc{\bfeta}{\mbox{\boldmath{$\eta$}}} 
\nc{\bx}{\bm{x}}
\nc{\bX}{\bm{X}}

\title{Dirichlet process mixture models for the Analysis of Repeated Attempt Designs}
\author[1]{Daniels, M.J.}
\author[2]{Lee, M.}
\author[3]{Feng, W.}
\affil[1] {mdaniels@stat.ufl.edu, Department of Statistics, University of Florida, Gainesville, FL}
\affil[2]{minjilee101@gmail.com, Edwards Lifesciences, Irvine, CA}
\affil[3]{weifeng.vivi@gmail.com, Keros Therapeutics, Lexington, MA}
\begin{document}

\maketitle

\noindent {\bf Abstract:}\\
In longitudinal studies, it is not uncommon to make multiple attempts to collect a measurement after baseline. Recording whether these attempts are successful provides useful information for the purposes of assessing missing data assumptions.  This is because measurements from subjects who provide the data after numerous failed attempts may differ from those who provide the measurement after fewer attempts.  Previous models for these designs were parametric and/or did not allow sensitivity analysis.  For the former, there are always concerns about model misspecification and for the latter, sensitivity analysis is essential when conducting inference in the presence of  missing data.  Here, we propose a new approach which minimizes issues with model misspecification by using Bayesian nonparametrics for the observed data distribution.  We also introduce a novel approach for identification and sensitivity analysis.  We re-analyze the repeated attempts data from a  clinical trial involving patients with severe mental illness and conduct simulations to better understand the properties of our approach.

\noindent {\bf Key Words:} Bayesian nonparametrics; informative priors; missing data 

\pagebreak

%{\bf update tables and other results from email on 3/6}

\section{Introduction}

In follow-up studies, multiple attempts are often made to collect a measurement after baseline (e.g., \citep{Wood2006}). We refer to these types of designs as {\em repeated attempt designs}.
The number of contact attempts made reflects the difficulty of obtaining the outcome data and
could provide additional information on the distribution of unobserved responses \citep{alho1990, LinSchaeffer95, Wood2006}. 
It is often the case that non-response or late response is related to the outcome data. Participants are more reluctant or less cooperative to answer questionnaires including sensitive questions, such as about income, alcohol consumption or smoking behavior.
A missing at random (MAR) assumption is likely too strong and not realistic in these studies.  Models for these designs are call {\em Repeated Attempt Models (RAM)}.

Most of the previous papers about repeated attempt designs exploited the information about multiple attempts in the context of selection models (SM) \citep{alho1990,Wood2006,jackson2010,jackson2012,chen:2018}.
One appealing aspect of selection models is that their factorization makes the parameters in the full data response model easily interpretable and usually these parameters are of primary interest.
However such models have limitations in that they are sensitive to model specification \citep{Kenward1998} and do not allow sensitivity analyses as in \cite{dh2008}, which are important for assessing unverifiable missing data assumptions. 

Pattern mixture models (PMMs) stratify subjects into patterns and model the response distribution as a mixture over patterns \citep{little1993,little1995}.
The correspondence between PMM's and the extrapolation factorization by PMMs makes them natural for sensitivity analyses \citep{dh2008} and have been proposed for repeated attempt designs \citep{daniels2015pattern}.
But when the number of patterns is relatively large, there will be some patterns with very few subjects and this leads to unidentified or weakly identified pattern-specific parameters \citep{little1995,dh2000}. 
Latent variable approaches \citep{roy2003,roy2008,lin2000,lin2004,muthen2003} introduce a discrete latent variable to discover and group the patterns in a data-dependent way and ad hoc approaches have also been used \citep{daniels2015pattern}; unfortunately the former rely on parametric specifications of the within 'class' distributions and the discrete latent variable for identifiability and are not robust to model misspecification. Bayesian nonparametric approaches can also be used to handle sparse patterns and avoid issues with model misspecification \citep{linero2015}.  We propose a Bayesian nonparametric approach here. This avoids the parametric assumptions used in approaches to collapse over dropout patterns in \citep{roy2003} and in previous RAMs.  In addition, we adapt the Bayesian nonparametric approach in \cite{linero2015} to a different setting (there, it was a longitudinal study with dropout) and also add covariates to the specification (which will also accomodate ignorable missingness in the covariates). 
We also introduce novel flexible priors for identification extending the previous development in \cite{daniels2015pattern}.

This work was motivated by the QUATRO trial \citep{gray2006}. The QUATRO trial was a single-blind, multi-center randomized controlled trial of the effectiveness of adherence therapy for schizophrenia. The trial was conducted in four centers and included 409 participants at baseline. Investigators then attempted to collect the  quality-of-life outcome at the end of follow-up of 52 weeks. As many as nine attempts were made but there were still some non-responders. 29 out of 204 (14\%) and 13 out of 205 (6\%) individuals failed to provide outcome data in the intervention and control groups respectively.   Due to the sparsity of subjects that had their outcome collected during 3 to 9 attempts,  those
subjects 
were merged into one pattern in the previous modeling \citep{daniels2015pattern}. 

Here, we propose a Dirichlet process mixture (DPM) model for the observed data.  In Section 2, we introduce the model for the observed data, priors for the observed data model parameters and novel priors for  the parameters of  the unidentified conditional distribution of the missing data given the observed data; this distribution has been called the {\em extrapolation distribution} \citep{dh2008}.  We re-analyze the QUATRO data using our model in Section~\ref{ch3sec:anal}. Simulation studies are presented in Section~\ref{ch3sec:simu}. We conclude with a discussion in Section~\ref{ch3sec:discuss}.

\section{Model}\label{ch3sec:model}
We propose a Dirichlet process mixture (DPM) model for the observed data distribution and priors for unidentified parameters in the extrapolation distribution based on 'corresponding' conditional distributions from the observed data. 

\subsection{Observed data model}

Denote the outcome as $Y$ and the vector of baseline covariates as $\bX$. $R$ will denote the number of attempts until the outcome is collected; there will be up to $K$ attempts in the study. Let $R=K+1$ correspond to the outcome not being obtained after the maximum number of attempts. Let $Z$ denote the intervention (1 for treatment and 0 for control).  We propose the following DPM,
\begin{align*}
Y_i | R_i, \bX_i,  Z_i; \bbeta_i, \sigma^2_i & \sim p(y | r, \bx, z; \bbeta_i, \sigma^2_i): r=1,,\ldots,K\\
R_i | \bX_i; \bxi_i  & \sim p(r | \bx; \bxi_i) \\
\bX_{i\ell} ; \bfeta_{ix} & \sim p(\bx_\ell ; \bfeta_{ix}), \ell=1,\ldots,p \quad \mbox{independent} \\
Z_i; \bfeta_{iz} & \sim p(z; \eta_{iz})\\
(\bbeta_i, \sigma^2_i, \bxi_i, \bfeta_i) | F & \sim F, \\
F & \sim DP ( \alpha, H_{0 \bbeta} \times H_{0 \sigma^2} \times H_{0 \bxi} \times H_{0 \bfeta} ),
\end{align*}
where $DP$ is a Dirichlet process prior and $\bfeta_i = (\bfeta_{ix}, \eta_{iz})$. The mass parameter $\alpha$ is given a Gamma prior and the base measures, $H_{\cdot}$ are specified as conjugate priors for the model parameters in $p(\cdot)$.  We provide details on the choice of distributions and priors in the context of the QUATRO data in Section 3.  

The joint distribution, $[Y, R, \bX, Z]$ can be expressed as an infinite mixture,

\begin{equation}\label{infinite:mixture}
f(y, r, \bx, z; \bbeta, \sigma^2, \bxi, \bfeta) = \sum_{j=1}^{\infty} \pi_j p(y | r, \bx, z; \bbeta_j, \sigma^2_j) p(r | \bx; \bxi_j) p(\bx; \bfeta_{jx}) p(z; \eta_{jz}),
\end{equation}
where 
    $\pi_j = \pi^\prime_j \prod_{s < j} (1-\pi^\prime_s)$  with 
    $\pi^\prime_j \sim Beta(1, \alpha)$. 
    
    The conditional distribution, $[Y | R, \bX, Z]$ can be expressed as follows:
$$p(y | {r}, {\bx}, z) = \sum_{j=1}^{\infty} w_j({r},{\bx}, z) p(y | {\bx}, {r}, z; {\bbeta}_j, \sigma^2_j),$$
  \noindent where 
  $$w_j({r}, {\bx}, z)=\frac{\pi_j  p({\bx}, z ; \bfeta_j) p({r}|{\bx, z};\bxi_j)}{\sum_{h=1}^\infty \pi_h  p({\bx}, z; \bfeta_{h}) p({r}|{\bx}, z;\bxi_h)}.$$
Clearly, the DPM provides a very flexible conditional distribution by default and will also provide a consistent estimator of the observed data distribution \citep{Ghos:1999}.
%where we assume a linear form for the mean function $\mu(z, \bm x, r) = \alpha^{(zr)} + \bm x\bm\bbeta_1$ and a constant variance over the number of attempts $\sigma^2(z, r)=\sigma^2$. More complex forms of the mean and variance are possible.

The DPM also accommodates ignorable missingness in covariates given the joint distribution of $[Y, R, \bX]$ is modelled;  i.e., it is a generative model.  Previous PMM and SM approaches do not allow missingness in covariates as they do not model $\bX$.

\subsection{Extrapolation Distribution}
The extrapolation distribution here is $[Y | R=K+1, \bX, Z]$.  This is not identified from the observed data as $Y$ is not observed when $R=K+1$.  We discuss identification in Section 2.4.

\subsection{Parameter of Interest}\label{ch3sub:theta}

The parameter of interest here is the marginal treatment effect unconditional on $R$ and $\bX$, 
\begin{align*}
\theta = & E(Y | Z=1) - E(Y | Z=0). 
\end{align*}
Each term on the right hand side can be computed from
\begin{equation}\label{param:int}
E(Y | Z=z) = \int \sum_{r=1}^{K+1} E(Y | Z=z, R=r, \bX=\bx)p(r | z,\bx) dF(\bx).
\end{equation}
Note in the above integral, we implicitly assume the distribution $F(\bx)$ does not depend on the intervention $Z$ (by randomization).  Also note given that our interest is in $\theta$, we only need to identify the {\em conditional expectation} for the extrapolation distribution, $E(Y | R=K+1, \bX, Z)$.  We discuss priors to identify this expectation (of the extrapolation distribution) in the next section.

\subsection{Priors for the Extrapolation Distribution}\label{ch3sub:priors}
\cite{daniels2015pattern} identified the above expectation assuming a monotone (linear) trend in $R$ on the expectation of $Y$ conditional on $Z$ and $\bX$, i.e., non-responders have worse outcomes than individuals who provide outcomes.  In the setting of many repeated attempts, like the QUATRO study ($K=9$), alternative specifications are needed as a linear specification is unlikely to be reasonable with many attempts.  In addition, the MAR assumption would imply $E(Y | R=K+1, \bX, Z) = E(Y | R < K+1, \bX, Z)$; this is likely not reasonable here (as it does not exploit the repeated attempt design).

For ease of notation, we set $\alpha^{(z,K+1)} = E(Y | R=K+1, \bX, Z)$, ignoring (for now) its dependence on $\bX$. 
The parameters $\alpha^{(z,K+1)}$ are not identified and we specify an informative prior for $\alpha^{(z, K+1)}$ conditional on the identified conditional means. We assume that non-responders in the missing pattern have worse outcomes than individuals who provide outcomes though our priors can be altered if this is not viewed as reasonable. Here we do not assume a parametric form for the expected value of $Y$ as a function of $R$ as in earlier work.

\noindent \textit{Triangular Priors}

We first consider two triangular priors.  Again, for ease of notation, we let $\alpha^{(zk)} = E(Y | R=k, \bX, Z)$.  The priors are
\begin{align*}
& \alpha^{(z, K+1)} | \alpha^{(z1)}, \cdots, \alpha^{(zK)} \sim T (\alpha^{(z)}_{min} - C^{(z)}, \:  \alpha^{(z)}_{min}, \: \alpha^{(z)}_{min} - C^{(z)}), \\
& \alpha^{(z, K+1)} | \alpha^{(z1)}, \cdots, \alpha^{(zK)} \sim T (\alpha^{(z)}_{min} - C^{(z)},  \: \alpha^{(z)}_{min}, \:\alpha^{(z)}_{min}), 
\end{align*}
where $T(a, b, c)$ is a triangular distribution with a lower limit $a$,  an upper limit $b$, and a mode $c$, $\alpha^{(z)}_{min} = \min(\alpha^{(z1)}, \cdots, \alpha^{(zK)})$ and
$C^{(z)} = (\alpha^{(z)}_{max} - \alpha^{(z)}_{min})\times P\%$.
$C^{(z)}$ specifies how far we want to move the `mean' parameter $\alpha^{(z,K+1)}$ from the minimum mean parameter for the observed outcomes.
$P$ is the sensitivity parameter and calibrates $C^{(z)}$ as percentage of the range of identified $\alpha^{(zd)}$. The first specification puts the most weight at the more extreme mean; the second puts the most weight at the minimum mean among the $R$ values corresponding to $Y$ being observed.  

\noindent \textit{Uniform Priors}

An alternative informative prior we consider for $\alpha^{(z, K+1)}$ is 
\begin{align*}
\alpha^{(z, K+1)} | \alpha^{(z1)}, \cdots, \alpha^{(zK)} \sim Unif(\alpha^{(z)}_{min}  - C^{(z)}, \alpha^{(z)}_{min} ),
\end{align*}
where $C^{(z)}$ is defined the same as in the triangular priors but here it quantifies the interval length of this uniform prior and describes how far we would move the lower bound of $\alpha^{(z,K+1)}$ from $\alpha^{(z)}_{min}$.  
$P$ is again the sensitivity parameter and calibrates $C^{(z)}$ as percentage of the range of identified $\alpha^{(zk)}$. 

\subsection{Computations} 
Posterior sampling and calculation of treatment effect $\theta$ in all models can be implemented using the R package \texttt{rjags} in R 3.3.2 software using a truncation approximation to the mixture model representation in (\ref{infinite:mixture}) (Ishwaran and James, 2001).\nocite{Ishw:Jame:2001} For generating triangular distributions, the R package \texttt{triangle} is used.   Details on how to do the MC integration in Section 2.3 can be found in the supplementary materials.

\section{Analysis of QUATRO}\label{ch3sec:anal}

\subsection{Models}

In the QUATRO trial, the outcome of interest $Y$ is the self-reported quality-of-life score and the set of baseline covariate $\bX$ includes the baseline score and the indicators of the 4 centers; the missingness in the baseline score can be accommodated in the DPM under an assumption of ignorable missingness. Up to 9 attempts were made to collect the 52 week outcome for participants ($K=9$). We set $Z = 1$ for treatment group and $Z = 0$ for control group. The patterns corresponding to 3 to 9 attempts in each treatment group   were very sparse, i.e., very few subjects in each of these patterns  (Table~\ref{ch3:tab1}). \cite{daniels2015pattern} merged those subjects into one pattern. There were an overall decreasing outcome mean with the number of attempts after merging and the MNAR assumption was viewed as plausible in that the unfavorable (lower) final quality-of-life score could be related to the late-responders. 
%{\bf Minji: clarify $Z$ here and how coded}

For our analysis, the number of attempts $R$ takes values in $\{1, 2, \cdots, 10\}$ and
$R=10$ corresponds to $Y$ missing even after all attempts.
The individuals with $Y$ missing but fewer than than the maximum number of attempts ($K$) were placed in the $R=10$ attempts pattern. 
We specify the DPM for the joint distribution of $(Y,R,\bX,Z)$ as follows
%{\bf tweak notation here so match with specification in section 2 - $\beta$, $\eta$ }
\begin{align*}
Y_i | R_i,  \bX_i,  Z_i; \bbeta_i, \alpha_i^{(zr)}, \sigma_i^{2} & \sim N(\alpha_i^{(zr)} +  \bX_i \bbeta_i, \sigma_i^{2}), \\
R_i |  \bX_i, Z_i; \bxi_i  & \sim \mbox{Multinomial} (1,  \bxi_i), \\
 \bX_{1i} | Z_i; \bfeta_i & \sim \mbox{Multinomial} (1,  \bfeta_i), \\
X_{2i} | m_{i},  \tau & \sim N(m_{i}, \tau_i^{2}), \\
Z_i & \sim \mbox{Ber}(p_i) \\
(\alpha_i, \bbeta_i,  \sigma_i^2, \bxi_i,  \bfeta_i, m_{i}, p_i) | F & \sim  F, \\
F & \sim DP ( \alpha, H_{0 \alpha^{(zr)}} \times H_{0 \bbeta}  \times H_{0 \sigma^2} \times H_{0 \bxi} \times H_{0 \bfeta}  \times H_{0 m} \times H_{0 \tau} \times H_{0 p}),
\end{align*}
where $\bX_{1i}$ is a vector of center indicators and $X_{2i}$ is the baseline score of $i$th observation, respectively. The base measures for the parameters are $H_{0 \alpha^{(zr)}} = N (\mu_{z}, \pi_{z}^2)$, $ H_{0 \bbeta} = N(\bbeta_0, \sigma_0^2)$, $H_{0 \sigma^2} = \mbox{Inv-Gamma} (a, b)$, $H_{0 \theta} = \mbox{Dirichlet}(\kappa_1)$, $H_{0 \bfeta} = \mbox{Dirichlet}(\kappa_2)$, $H_{0 m} = N(m, \tau^2)$, $H_{0 \tau} = \mbox{Inv-Gamma} (c, d)$, and $H_{0 p} = \mbox{Beta}(e, f)$.
%For notational clarity, we did not follow the specific
We follow   \cite{linero2015} and Roy et al. (2018)\nocite{Roy:2018} for specification of the hyperparameters of the distributions $H_{0 \cdot}$ (details in the supporting information). 
%{\bf Minji: need these details in the supp materials including As part of the prior specification, the data was standardized with mean 0 and variance 0.5. - working on with other stuff in different tex file.}

% {\bf Minji: need to specify what all the H are.}
% {\bf Minji: does mean of $X_{2i}$ depend on $Z$?}

We considered models with the conditional distribution of the outcome in the DPM depending in the  full number of attempts, $R$ ($K = 9$) and on the merged attempts, collapsing $K\in \{3,\ldots,9\}$ into one group as in \cite{daniels2015pattern}. For the model with conditional depending on merged attempts, let $R^{\star}$ denote the number of (merged) attempts and the individuals with missing outcome were placed in $R^{\star} = 4$. For the merged attempt specification, we changed the form of $[Y_i | R_i,  \bX_i,  Z_i; \bbeta_i, \alpha_i^{(zr^{^{\star}})}, \sigma_i^{2}]$ to $N(\alpha_i^{(zr^{^{\star}})} +  \bX_i \bbeta_i, \sigma_i^{2})$; so now the intercepts depend on $R^\star$ instead of $R$.  But note we still model $[R | \bX, Z]$.
%$R^{\star}_i |  X_i, Z_i; \theta^{\star}_i  \sim \mbox{Multinomial} (1,  \theta^{\star}_i)$. 

%{\bf Minji: also need to specify $\alpha_i^{(zr)}$; don't we consider where this varies by $R$ and by $R^\star$?  please provide more details on what models we considered and fit.  we could discuss the different forms for $Y_i | \cdots$ in the text.  we might want to compare the fits using LPML - R package by Vehtari and Gelman to do this} 

It is not uncommon that individuals who do not respond do not receive the maximum number of attempts. Here we assign them to the $K+1$ group.  However, they could be viewed as censored.  
We discuss this issue of censored $R$ in Section~\ref{ch3sec:discuss}.

\subsection{Results and Sensitivity Analysis}
We computed the expected log pointwise predictive density via R package \texttt{loo} for the model with the intercept depending on $R$ and depending on $R^\star$. The point estimates (standard errors) for model with the conditional depending on the full number of attempts ($R$) and the one with conditional depending on the merged attempts ($R^\star$) were -2258.4 (65.8) and -2117.6 (33.9).  Thus there was some evidence for using the 'simpler' conditional in the DPM. 

The marginal treatment effects as defined in Section 2.3  from different model and prior specifications are displayed in Table~\ref{ch3:datatab1}.  The sensitivity parameter for the RAM-PMM, denoted by $C$ was set to $3$ as in \cite{daniels2015pattern}, which corresponds to those subjects who never provided the outcome being assumed the same as those in the last (i.e., the third) collapsed pattern.  
Prior specifications for the extrapolation distribution are as introduced in Section 2.4.  
We 
let {\em unif$_{P}$} denote the uniform prior with sensitivity parameter $P$. The triangular prior with mode $\alpha^{(z)}_{min} - C^{(z)}$ and sensitivity parameter $P$ is denoted by {\em tri1$_{P}$} and the one with mode $\alpha^{(z)}_{min}$ and $P$ is denoted by tri2$_{P}$. Here we considered $P = 10$ and 20.  {\em Point mass (pm)} is the point mass prior at $\alpha^{(z)}_{min}$. 

The treatment effects were similar across all priors and models, all supporting a negative treatment effect but with 95\% credible intervals all covering zero;   the lack of differences between the priors was related to the low proportion of missingness (we explore this in the simulations).   The CIs for the non-point mass priors were all slightly wider than the point mass prior.  And the treatment effect under MAR was attenuated.  Conclusions did not differ when using the PMM or the selection model though the point estimate from the selection model was farthest from zero.

%{\bf Discuss results in table when completed including lengths of CIs - Please give some times to check the performance of model without $K+1$. I realized that I needed to re-run the data analysis. }

\subsection{Goodness of Fit}
To assess goodness of fit of the outcome distribution, we compute the observed data means from the DPM as follows
\begin{equation}\label{param:int2}
\int \sum_{r=1}^{K} E(Y | Z=z, R=r, \bX=\bx) p(r | z, \bx) dF(\bx|z).
\end{equation}
 The difference from (\ref{param:int}) is the sum is only up to $K$ and we now integrate over the distribution of $[\bX | Z]$. 
The estimated expectations of the response given each merged attempt are recorded in Table \ref{ch3:datatab2} in the supporting information for both the conditional depending on $R$ (full) and depending on $R^\star$ (merged). The observed data means are very similar to the means from the DPM and the 95\% CIs covered the observed data means in all cases.  So the fit of the DPM seemed good.

\section{Simulation Study}\label{ch3sec:simu}

We conducted a simulation study designed to better understand the estimation of the treatment effect using the proposed approach versus alternative parametric approaches.

\subsection{Design}

We investigated several data generating scenarios. In scenario (1), we considered datasets with a truth based on the estimates from QUATRO data analysis result, but with two different sample sizes, a sample size similar to QUATRO ($409$) and a larger sample size ($1000$). In the other scenarios datasets were generated with sample sizes $N = 500$ and $N = 1000$ using different models and different functional forms for $E[Y | R, Z]$.  We consider several different error distributions for all these scenarios. For the distribution of the number of attempts, we computed the probability of the number of attempts after attempt 3 and divided it equally to compute the  probabilities for attempts 4 - 9; the probability of $R=K+1=10$ was based on the QUATRO data.  And these probabilities were used to generate the distribution of $R$. Outcome data obtained at $R=K+1$ attempts were set as missing.    We also considered more missingness than we observed in QUATRO for two of the scenarios.   
%{\bf ****}

%{\bf Minji: don't we also have a Scenario (4) - simulate under SM based on Quatro?  and i wonder if we want a Scenario (5) which would allow a nonlinear relationship with x (instead of linear).  this would likely be a case where the SM and PMM would do poorly but the proposed approach would do well.}

%{\bf also need to do (might be doing now): simulated under SM; fit PMM without collapsing}

For all scenarios, we only considered one covariate $X$ and assumed the coefficient for covariate is constant across groups. For first four scenarios, we assumed three different distributions for $Y | X=x, R=r, Z=z$: normal distribution, student's t distribution with three degrees of freedom and skew normal distribution with skewness parameter $\alpha = 3$ \citep{azzalini_2013}.  For the first three scenarios, we specified $ \alpha^{(zr)} + x\beta_1$ as the location parameter and $\sigma^2$ as the scale parameter. For generating from the skew normal distribution, the R package \texttt{sn} was used. 
%{\bf minji: need a reference for your parameterization of skew normal}

For all scenarios, we fitted the DPM with the prior specifications considered in the data example and the PMM-RAM models with merged attempts described in Section \ref{ch3sec:anal} (and sensitivity parameter set to $C=3$) except scenario $(4)$. In that scenario, the PMM-RAM without collapsing attempts was also fitted. We fit the selection model (SM)  for scenarios (4) and (5).

%{\bf Minji: need sim details in supp materials}

Further details on the simulation scenarios for the observed data are as follows.
\begin{itemize}
\item[(1)]  For this scenario, to generate the replicated data, we used the QUATRO data and fitted conditional distribution of the outcome for the PMM-RAM with subjects with 3 to 9 attempts merged and baseline scores as a covariate. %{\bf For the distribution of the number of attempts, we computed the probability of the number of attempts after attempt 3 and divided it equally to compute the  probabilities for attempts 4 - 9; the probability of $R=K+1=10$ was based on the QUATRO data.  And these probabilities were used to generate the distribution of $R$. Outcome data obtained at $R=K+1$ attempts were set as missing.}   {\bf is the preceding text in bold true for all scenarios except for Scenario 4?  if so, move to **** above}

\item[(2)] 
For this scenario (and for scenario 3) we again fit a PMM-RAM but assuming that $\alpha^{(zr)}$ is a deterministic function of $(z, r)$, $z = 0, 1$, $r = 1, \ldots, R$. We set the functional form $h^*$ to be linear as follows, 
\begin{align*}
h^\star(z, r^\star) = z(27.24 - 1.91 r^\star) + (1 - z)(25.58 - 1.65 r^\star),
\end{align*}
where $r^\star = 1, \ldots, 3$. We assume $\alpha^{(z,r)}=h^\star(z, 3)$ for $r = 4,\ldots,9$. For this scenario (and scenario 3), the one covariate $X$ was generated from a normal distribution $N(2, 0.2)$. 
We set the coefficient $\beta_1 = 0.4$.

\item[(3)] This was the same as scenario (2) but now with a nonlinear functional form $h$ as follows:
\begin{align*}
h(z, r) = z(30\exp(-0.13r)) + (1 - z)(29\exp(-0.15r)), 
\end{align*}
where $r = 1, \ldots, 9$. 

\item[(4)] We fitted the SM for the QUATRO data with baseline score via the \textit{Stata} module \textit{alho}. Since this module requires each attempt have at least one observation, we used 8 attempts and attempt 9 was treated as missing. Then the outcome and number of attempts were generated sequentially from 
\begin{equation*}
Y | X = x, Z = z  \sim N( \beta^{\star}_0 + \xi z + x \beta^{\star}, \sigma^{\star}),
\end{equation*}
\begin{equation*}
P(R = r | R \geq r, Z = z, X = x, Y = y)  = \mbox{expit}(\lambda_{0r} + \gamma_{r}z + \lambda_{r}x + \delta_1 y + \delta_2 y z),
\end{equation*}  
where $\beta^{\star}$, $\xi$, $\sigma^{\star}$, $\lambda_{0r}$, $\gamma_{r}$, $\lambda_{r}$, $\delta_1$ and $\delta_2$ were estimated from the QUATRO data.  

As mentioned above, to run the \textit{alho} module, at least one observation is needed for each number of attempts. This is not a restriction for the DPM.  As a result, we considered comparisons using only those datasets that met the restriction and also using all the datasets (but without the SM). 
Note that 49\% of the datasets had no outcomes for at least one attempt at a sample size of  500.  

\item[(5)] We considered a complex two component mixture model with one component with a linear functional form for the mean ($M=1$) and the other with a nonlinear functional form for the mean ($M=2$) as follows,
\begin{align*} 
&Y | X=x, R=r, Z = z, M = 1 \sim N (g(z, r) + x\beta_1, \sigma^2), \\
&Y | X=x, R=r, Z = z, M = 2 \sim N(g^{\star}(z, r) + x\beta_2, \sigma^2), 
\end{align*}
where
$g(z, r) = (60.24 - 1.91r)z + (60.58 - 1.65r)(1-z)$,
$g^{\star}(z, r) = z(30\exp(-0.13r)) + (1 - z)(29\exp(-0.15r))$,
$\beta_1 = 0.4$, $\beta_2 = 1$, and $\sigma = 2$.  The parameter values were similar to the other scenarios based on QUATRO. 

We considered weights:
$$\pi_{M = 1} (Z, R) = \frac{\exp(2 Z -  0.2  R - 1)}{1 +  \exp(2  Z -  0.2 R - 1)}, \quad \pi_{M = 2} = 1 - \pi_{M = 1}.$$
%and
%$$\pi_{M = 1} (Z, R^{\star}) = \exp(2 Z - 0.1ZR^{\star} - 0.1 R^{\star} - 1) / (1 +  \exp(2 Z - %0.1ZR^{\star}  - 0.1 R^{\star}  - 1)).$$  

\item[(6)]   This scenario was explicitly a latent class model (LCM). We assumed the outcome intercept, $\alpha^{(zc)}$ is a deterministic function of treatment ($Z=z$) and latent class ($C=c$), where $c = 1, \ldots, 4$.   We set the functional form $h^*$ to be linear as follows, 
\begin{align*}
h^\star(z, c) = z(27.24 - 1.91 c) + (1 - z)(25.58 - 1.65 c).
\end{align*}
The classes, $C$ and the number of attempts, $R$ are connected as follows:
$pr(C = 1 | R = 1) = pr(C = 2 | R = 2) = pr(C = 3 | 3 \leq R \leq 9) = 0.8$, $pr(C = 2 | R = 1) = pr(C = 3 | R = 1)$ $= pr(C = 1 | R = 2) = pr(C = 3 | R = 2) =$ $pr(C = 1 | R = 3) = pr(C = 2 | R = 3) = 0.1$ and $pr(C = 4 | R = 10) = 1$.
As in some of the previous scenarios, the one covariate $X$ was generated from a normal distribution $N(2, 0.2)$. 
We set the coefficient $\beta_1 = 0.4$.

\end{itemize}

The unidentified location parameter in the extrapolation distribution,  $\alpha^{(z, K+1)}$ was specified as follows for each scenario, (1) $C = 3$ (same as last identified pattern); (2) $h^\star(z,4)$ (one unit past last identified collapsed pattern); (3) $h(z,10)$ (one unit past last identified pattern); (4) NA (implicitly identifed in parametric SM); (5) $E(Y | X, R=10, Z)$; (6) NA (defined in the specification).

One thousand datasets were simulated for each scenario. Bias $\frac{1}{1000} \sum_{k=1}^{1000}(\hat{\theta}_k - \theta)$, mean squared error (MSE) 
$ \frac{1}{1000}\sum^{1000}_{k=1} (\hat{\theta}^{(k)} - \theta)^2$,
and empirical coverage probability (and length) of $95\%$ (equal tail) credible intervals were computed to assess the estimation of $\theta$, where $\theta$ is the true value for marginal treatment effect and $\hat{\theta}^{(k)}$ is the estimate in $k$th simulated dataset.

\subsection{Results}
We summarize the results for each scenario below.  For each scenario, the DPM 'none' column (in the tables) illustrates the robustness of the DPM for any observed data distribution (ignoring the $K+1$ pattern, i.e., the extrapolation distribution); the other rows illustrate the impact of the sensitivity approach.

\noindent {\em Scenarios (1)-(3)}

For scenario (1), the results for the DPM under different prior specifications showed close to the nominal coverage, and bias and MSE smaller for the collapsed pattern data.  The results were also robust to different error distribution specifications (as expected).  See Tables~\ref{scen1:DPM:normal},  \ref{sup:scen1:DPM:tdist}, \ref{sup:scen1:DPM:skewed}, \ref{sup:scen1:PMM:collapsed}, and \ref{sup:scen1:PMM:SM} in the supporting information.

The PMM-RAM could only be fit on the collapsed pattern data.  The results for the normal errors were similar to the DPM but for the t- and skew normal errors, the MSEs were larger; however, the coverage was still close to the nominal level. The selection model (SM) also had larger MSEs for the skew normal errors.  

The results for scenarios (2) and (3) mirrored the key results from scenario (1). See Tables~\ref{scen2:DPM:skewed} and \ref{scen2:PMM:SM} and Tables~\ref{sup:scen2:DPM:normal}, \ref{sup:scen2:DPM:tdist}, \ref{sup:scen2:PMM:collapsed}, \ref{sup:scen3:DPM:normal}, \ref{sup:scen3:DPM:tdist}, \ref{sup:scen3:DPM:skewed}, \ref{sup:scen3:PMM} and \ref{sup:scen3:PMM:SM} in the supporting information.

\noindent {\em Scenario (4)}

Under the selection model truth, coverage for the DPM was close to the nominal level under all priors and the results in terms of MSE were very similar to the SM fit.  See Tables~\ref{scen4:DPM}, \ref{sup:scen4:PMM} and \ref{sup:scen4:SM} in the supplementary material.

\noindent {\em Scenario (5)}

The DPM had much smaller MSE and bias than the PMM-RAM and SM as expected as well as close to the nominal coverage; the coverage of the PMM-RAM and SM decreased with sample size going as low as $.87$;
see Tables~\ref{scen5:DPM} and \ref{scen5:PMM:SM}.

\noindent {\em Scenario (6)}

The DPM does a good job in this scenario with low bias and coverage a bit over the nominal level for both sample sizes; see Tables~\ref{sup:scen6:N=500} and \ref{sup:scen6:N=1000} in the supporting information.

\noindent {\em Exploration of different amounts of missingness}

We explored the impact of varying amounts of missingness, $10, 25, 35 \%$, for 
scenarios (2) and (6); see Tables \ref{sup:scen6:N=500}-\ref{sup:scen2:N=1000} in the supporting information.   As expected, the length of credible intervals increases with the amount of missingness.  In addition, the impact of the different prior specifications is larger when the amount of missingness increases.  We also note that the difference in inference on $\theta$ between the priors is a function of the amount of missingness {\em and} the range (i.e., $\alpha_{max}-\alpha_{min}$) and sensitivity parameter $P$.    
%{\bf put in response?} also comment somewhere that bias not necessarily go to zero (but coverage stays close to nominal)

\section{Discussion}\label{ch3sec:discuss}

We have proposed a Bayesian nonparametric approach, using Dirichlet process mixtures, for repeated attempt designs. It is not uncommon to observe sparse patterns. Instead of merging sparse patterns in an ad hoc way as in \cite{daniels2015pattern}, this approach implicitly deals with sparse patterns in a data-dependent way, allows for (ignorable) missingness in covariates, and allows for sensitivity analysis.
However, the conclusions here were not substantively different from those reported in
\cite{daniels2015pattern} due to low missingness.  

To investigate the performance of our DPM models,   including in settings with more missingness than QUATRO,   we conducted a simulation study in Section~\ref{ch3sec:simu}.  The DPM approach resulted in desirable frequentist operating characteristics (bias, MSE, coverage) across all scenarios.  The approach to sensitivity analysis using different priors for the mean of the extrapolation distribution also resulted in good frequentist properties.  The PMM-RAM and SM worked quite well in all scenarios except for the complex distribution in the last scenario.  However, we also point out that the PMM-RAM could not be fit to settings with a large number of sparse patterns (we collapsed in the simulations) and the SM also had issues with sparse patterns as the macro requires at least one observation in each pattern.  
%Overall the latent pattern mixture model with the correct number of classes had the highest posterior probability and the model averaging approach performed reasonably well in estimating the marginal intervention effect $\theta$. 

There are numerous extensions to our approach. 
In the analysis of QUATRO data, individuals who failed to provide outcome data were all placed in the $R=K+1$ attempt group, regardless of the number of attempts made to obtain these outcomes. If we treated these as censored, they were assigned to 'patterns' based only on $[R, X]$ which here resulted in most {\em not} being in the $K+1$ group; we viewed this as unreasonable in this application.  

We assumed that that individuals who do not respond have lower scores as the number of contact attempts increases. For other analyses, it is possible that those without an observed response are more similar to those observed after a few contact attempts than after many contact attempts; this could be addressed by specifying alternative priors for the extrapolation distribution expectation.
To further improve estimation of the necessary conditional expectations, $E(Y | R, \bX, Z)$ the DPM can be replaced by an enriched DPM (Wade et al., 2011).\nocite{Wade:2011}    It would also be of interest to extend the current approach to repeated attempt settings with multiple modalities as in M{\"u}ssner et al. (2015).\nocite{Muss:2015}  

\section*{Acknowledgments}
Partially funded by NIH R01 CA183854, HL 158963, and HL 166324. %{The majority of the work was done when Minji Lee was at the University of Florida.} 

\section*{Supporting information}
Web appendices and Tables referenced in Sections 3 and 4.2 are available with this paper at the Biometrics website on Wiley Online Library.
%The supplementary materials contains additional details on computations from Section 3 and additional simulation results from Section 4.2.

\bibliographystyle{apa}
\bibliography{reference}

\pagebreak
\begin{table}[ht]
\centering
\renewcommand{\arraystretch}{1.5}
\scriptsize
\caption{QUATRO data: counts (outcome means) by  number of attempts ($k$) and  randomized group ($Z$).} 
\label{ch3:tab1}
\begin{tabular}{@{}lccccccccc@{}c@{}}
\hline & \multicolumn{9}{c}{ $Y$ observed after $k$ attempts} & $Y$ missing\\
\# of attempts ($k$) &          1 &          2 &             3 &          4 &          5 &          6 &          7 &          8 &     9 &               \\
\hline
       Control ($n=205$) &     77(42.4) &     94(41.3) &   7(38.7) &    7(34.7) &     3(34.2) &   2(32.9) &    1(40.7) &   1(62.98) &    0(NA) &   13  \\

     Treatment ($n=204$) &     73(40.7) &     90(40.2) &   7(38.6) &    1(45.7) &   3(35.0) &     0(NA) &      0(NA) &    1(30.3) &    0(NA) &   29 \\
\hline
\end{tabular}

\end{table}

\begin{table}[htpt]
\centering
\renewcommand{\arraystretch}{1.5}
\footnotesize
\caption{Results: Posterior summaries for treatment effect $\theta$ for different prior and model specifications.  Full corresponds to modelling the full number of attempts ($K=9$) (i.e., $Y|R$) and merged corresponds to modeling the merged number of attempt $K=3$ (i.e.,  replacing $R$ with $R^\star$) in the DPM given in Section 3.1;  full with merged corresponds to the DPM with the conditional only depending on the merged attempts (i.e., $Y|R^\star$) but still modelling $R$.
  The first eight rows correspond to DPM specifications with different priors for the extrapolation distribution conditional mean.  PMM and SM correspond to the pattern mixture models and selection models from earlier work. MAR corresponds to ignorable MAR.}
\label{ch3:datatab1}

\begin{tabular}{@{}ccccccc@{}}
\hline
& \multicolumn{2}{c}{Full}& \multicolumn{2}{c}{Full with merged conditional} &  \multicolumn{2}{c}{Merged} \\
\hline
model & mean   &   95\% CI  / length of CI & mean   &   95\% CI  / length of CI  &   mean   &   95\% CI / length of CI   \\
\hline
completer & -0.44 & (-2.39, 1.61) / 4.00 & -0.44 & (-2.47, 1.58) / 4.05 & -0.48 & (-2.59, 1.57) / 4.16 \\

point mass & -0.43 & (-2.30, 1.48) / 3.78 & -0.39 & (-2.37, 1.69) / 4.06 & -0.48 & (-2.69, 1.75) / 4.44 \\

unif$_{10}$ & -0.43 & (-2.31, 1.48) / 3.79 & -0.39 & (-2.38, 1.70) / 4.08 & -0.48 & (-2.69, 1.75) / 4.44 \\

unif$_{20}$ & -0.43 & (-2.31, 1.48) / 3.79 & -0.39 & (-2.38, 1.70) / 4.08 & -0.48 & (-2.69, 1.76) / 4.44 \\

tri1$_{10}$ & -0.43 & (-2.31, 1.48) / 3.79 & -0.39 & (-2.38, 1.70) / 4.08 & -0.48 & (-2.69, 1.75) / 4.44 \\

tri1$_{20}$ & -0.43 & (-2.31, 1.48) / 3.79 & -0.39 & (-2.38, 1.70) / 4.08 & -0.48 & (-2.69, 1.75) / 4.44 \\

tri2$_{10}$ & -0.43 & (-2.31, 1.48) / 3.79 & -0.39 & (-2.38, 1.69) / 4.07 & -0.48 & (-2.69, 1.75) / 4.44 \\

tri2$_{20}$ &  -0.43 & (-2.31, 1.47) / 3.79 & -0.39 & (-2.38, 1.70) / 4.08 & -0.48 & (-2.69, 1.76) / 4.44 \\

PMM & & & & & -0.7 & (-3.5, 2.0) / 5.5 \\

SM & -1.6 & (-3.9, 0.7) / 4.6 & & &-1.5 & (-3.9, 0.8) / 4.7 \\

MAR & & & & & -0.50 & (-3.19, 2.04) / 5.23 \\

\hline

\end{tabular}
\end{table}

\begin{table}[htpt]
\centering
\renewcommand{\arraystretch}{1.5}
\footnotesize
\caption{Scenario 2 with skewed normal distribution: Bias, MSE and interval coverage probability and length for the estimated treatment effect $\theta$ (MC standard error in parentheses), based on 1000 replications. The sample size is 500 and 1000, respectively. The subscript 4 represents that subjects with 3 to 9 attempts were merged to 3 attempts ($K = 3$). Prior equal to 'none' corresponds to the estimate of the treatment effect without the missing group $K+1$ (so completers only). The notation for the different priors can be found in Section 3.2. }
\label{scen2:DPM:skewed}

\begin{tabular}{@{}ccccccccccccc@{}}
%1000_4_no
\hline
Scenario &  \multicolumn{3}{c}{ 500$_4$ } & \multicolumn{3}{c}{ 500 }   &  \multicolumn{3}{c}{ 1000$_4$ } & \multicolumn{3}{c}{ 1000 } \\

Prior & bias  &  MSE &  coverage & bias  &  MSE  &  coverage & bias  &  MSE  &  coverage & bias  &  MSE  &  coverage\\
\hline

None & -0.010 & 0.358 & 0.956  &  -0.009 & 0.361 & 0.959 & -0.021 & 0.193 & 0.949 & -0.019 & 0.191 & 0.948 \\

% & (0.02) & & (2.394) & (0.02) & & (2.401) & (0.01) & & (1.713) & (0.01) & & (1.705) \\

p.m & -0.008 & 0.359 & 0.960 & -0.011 & 0.368 & 0.954 & -0.020 & 0.195 & 0.949 & -0.023 & 0.191 & 0.947 \\

% & (0.02) & & (2.405) & (0.02) & & (2.429) & (0.01) & & (1.719) & (0.01) & & (1.730) \\
 
unif$_{10}$  & -0.009 & 0.359 & 0.959 & -0.012 & 0.369 & 0.955 & -0.022 & 0.196 & 0.948 & -0.024 & 0.192 & 0.948 \\

% & (0.02) & & (2.408) & (0.02) & & (2.434) & (0.01) & & (1.721) & (0.01) & & (1.734) \\

unif$_{20}$  & -0.010 & 0.360 & 0.959 & -0.014 & 0.369 & 0.954 & -0.023 & 0.196 & 0.949 & -0.026 & 0.192 & 0.949 \\

% & (0.02) & & (2.410) & (0.02) & & (2.439) & (0.01) & & (1.724) & (0.01) & & (1.739) \\

tri1$_{10}$  & -0.009 & 0.360 & 0.959 & -0.013 & 0.369 & 0.955 & -0.022 & 0.196 & 0.948 & -0.025 & 0.192 & 0.948 \\

% & (0.02) & & (2.408) & (0.02) & & (2.435) & (0.01) & & (1.722) & (0.01) & & (1.736) \\

tri1$_{20}$  & -0.011 & 0.360 & 0.959 & -0.014 & 0.370 & 0.954 & -0.024 & 0.196 & 0.948 & -0.027 & 0.193 & 0.948 \\

% & (0.02) & & (2.412) & (0.02) & & (2.442) & (0.01) & & (1.726) & (0.01) & & (1.742) \\

tri2$_{10}$  & -0.008 & 0.359 & 0.959 & -0.012 & 0.368 & 0.954 & -0.021 & 0.195 & 0.948 & -0.024 & 0.192 & 0.947 \\

% & (0.02) & & (2.407) & (0.02) & & (2.432) & (0.01) & & (1.721) & (0.01) & & (1.733) \\

tri2$_{20}$  & -0.009 & 0.360 & 0.959 & -0.013 & 0.369 & 0.955 & -0.022 & 0.196 & 0.948 & -0.025 & 0.192 & 0.949  \\

% & (0.02) & & (2.408) & (0.02) & & (2.435) & (0.01) & & (1.722) & (0.01) & & (1.736) \\
\hline

\end{tabular}
\end{table}

\begin{table}[htpt]
\centering
\renewcommand{\arraystretch}{1.5}
\footnotesize
\caption{Scenario 2 with skewed normal distribution for PMM-RAM and selection model (SM): Bias, MSE and coverage probability (and interval length) for the estimated treatment effect $\theta$ (MC standard error in parentheses), based on 1000 replications. The sample size is 500 and 1000, respectively. The sensitivity parameter is set to $C = 3$ for the RAM-PMM.}
\label{scen2:PMM:SM}

\begin{tabular}{@{}ccccccc@{}}

\hline
Scenario &  \multicolumn{3}{c}{ 500 } &  \multicolumn{3}{c}{ 1000 } \\

 & bias &  MSE &  coverage & bias &  MSE  &  coverage \\
\hline

PMM & -0.015 & 0.409 & 0.959 & -0.025 & 0.206 & 0.961 \\

% & (0.02) & & (2.599)  & (0.01) & & (1.803) \\
 
SM & -0.020 & 0.453 & 0.948 & -0.035 & 0.235 & 0.941 \\

% & (0.02) & & (2.600)  & (0.02) & & (1.834) \\

\hline

\end{tabular}
\end{table}

% working on 

\begin{table}[htpt]
\centering
\renewcommand{\arraystretch}{1.5}
\footnotesize
\caption{Scenario 5: Bias, MSE and interval coverage probability and length for the estimated treatment effect $\theta$ (MC standard error in parentheses), based on 1000 replications. The sample size is 500 and 1000, respectively. Prior equal to 'none' corresponds to the estimate of the treatment effect without the missing group $K+1$ (so completers only). The notation for the different priors can be found in Section 3.2. }
\label{scen5:DPM}

\begin{tabular}{@{}ccccccc@{}}

\hline
Scenario &   \multicolumn{3}{c}{ 500 } & \multicolumn{3}{c}{ 1000 }  \\

Prior &  bias  &  MSE  &  coverage & bias  &  MSE  &  coverage \\
\hline

None &  -0.001 & 0.097 & 0.954 & 0.012 & 0.045 & 0.969  \\

% & (0.01) & & (1.265) & (0.01) & & (0.914) \\

p.m &  -0.043 & 0.103 & 0.955 & -0.015 & 0.048 & 0.972 \\
 
% &  (0.01) & & (1.323) & (0.01) & & (0.957)  \\
 
unif$_{10}$  & -0.032 & 0.103 & 0.952 & -0.003 & 0.048 & 0.973\\

%&(0.01) & & (1.331) & (0.01) & & (0.963)   \\

unif$_{20}$  & -0.020 & 0.103 & 0.954 & 0.009 & 0.049 & 0.976\\

% &(0.01) & & (1.339) & (0.01) & & (0.969)  \\

tri1$_{10}$  &  -0.028 & 0.103 & 0.952 & 0.001 & 0.048 & 0.973 \\

%& (0.01) & & (1.333) & (0.01) & & (0.965)   \\

tri1$_{20}$  &  -0.013 & 0.103 & 0.956 & 0.016 & 0.049 & 0.975 \\

%& (0.01) & & (1.345) & (0.01)  & & (0.974)   \\

tri2$_{10}$  &  -0.035 & 0.103 & 0.951 & -0.007 & 0.048 & 0.972\\

%& (0.01) & & (1.328) & (0.01) & & (0.961)   \\

tri2$_{20}$  & -0.028 & 0.103 & 0.953 & 0.001 & 0.048 & 0.972 \\

%& (0.01) & & (1.333) & (0.01) & & (0.965)  \\
\hline

\end{tabular}
\end{table}

\begin{table}[htpt]
\centering
\renewcommand{\arraystretch}{1.5}
\footnotesize
\caption{Scenario 5 with PMM-RAM and selection model (SM): Bias, MSE and coverage probability (and interval length) for the estimated treatment effect $\theta$ (MC standard error in parentheses), based on 1000 replications. The sample size is 500 and 1000, respectively. The sensitivity parameter is set to $C = 3$ for the RAM-PMM.}
\label{scen5:PMM:SM}

\begin{tabular}{@{}ccccccc@{}}

\hline
Scenario &  \multicolumn{3}{c}{ 500 } &  \multicolumn{3}{c}{ 1000 } \\

 & bias &  MSE &  coverage & bias &  MSE  &  coverage \\
\hline

PMM & -0.155 & 0.134 & 0.928 & -0.136 & 0.070 & 0.915 \\

% & (0.01) & & (1.315)  & (0.01) & & (0.915) \\
 
SM & -0.233 & 0.182 & 0.897 & -0.211 & 0.104 & 0.868 \\

% & (0.01) & & (1.378)  & (0.01) & & (0.973) \\

\hline

\end{tabular}
\end{table}

\end{document}

\title{Supporting Information for  ``Dirichlet process mixture models for the Analysis of Repeated Attempt Designs with Sparse Patterns''}
\date{}
\maketitle

%{\bf REMAINING - REMOVE MC ERROR ROWS -WAIT IN CASE ANY UPDATED SIMULATION TABLES}

\section{Computations}
We obtained  posterior samples using the R package \texttt{rjags} using the blocked Gibbs sampler approach \citep{Ishw:Jame:2001}.  We sampled 50000 iterations with a burn-in of 5000 iterations. To calculate the treatment effect $\theta$ (see Section 2.3), we took every fifth posterior samples and computed $E (Y | Z = z)$ based on
\begin{align*}
E(Y | Z = z) & = \int \sum_{r = 1}^{K + 1} E[Y | Z = z, R = r, \bX = \bx] p(r | z, \bx) dF(\bx)\\
&=  \int \sum_{r = 1}^{K + 1} \int \sum_{j = 1}^{H} w_j (r, \bx, z) \: y \: p(y | \bx, r, z; \bbeta_j) dy \; p(r | z, \bx) \; dF(\bx) \\
& = \int \sum_{r = 1}^{K} \int \sum_{j = 1}^{H} w_j (r, \bx, z) \: y \: p(y | \bx, r, z; \bbeta_j) dy \; p(r | z, \bx) \; dF(\bx)  \\
& \qquad +  \int  E[Y | Z = z, R = K + 1, \bX = \bx]\;  p(R = K + 1 | z, \bx) \; dF(\bx) \\
& = \int \sum_{r = 1}^{K} \sum_{j = 1}^{H} w_j (r, \bx, z) ( \alpha^{(zr, j)} + \bx \bbeta^{(j)} ) \; p(r | z, \bx) \; dF(\bx) \\
& \qquad +  \int  E[Y | Z = z, R = K + 1, \bX = \bx] p(R = K + 1 | z, \bx) \; p (\bx) \; d\bx \\
& = \int \sum_{r = 1}^{K} \sum_{j = 1}^{H} w_j (r, \bx, z) ( \alpha^{(zr, j)} + \bx \bbeta^{(j)} ) \sum_{m = 1}^{H} v_m (\bx, z) 
p (r; \bxi_m) \; p (\bx)\; d\bx\\
& \qquad +  \int  \alpha^{(z,K+1)}(\bx) 
\; \sum_{m = 1}^{H} v_m (\bx, z)\; 
p (R = K + 1; \bxi_m) \; p (\bx)\;  d\bx \\
& = \int \sum_{r = 1}^{K} \sum_{j = 1}^{H} w_j (r, \bx, z) ( \alpha^{(zr, j)} + \bx \bbeta^{(j)} ) \;  \sum_{m = 1}^{H} v_m (\bx, z) 
p (r; \bxi_m) \sum_{l = 1}^{H} \pi_l \: p (\bx; \bfeta_l) \; d\bx \\
& \qquad +  \int \alpha^{(z,K+1)}(\bx) \; \sum_{m = 1}^{H} \; v_m (\bx, z)\; 
p (R = K + 1; \bxi_m) \sum_{l = 1}^{H} \pi_l \: p (\bx; \bfeta_l) d\bx, \\
\end{align*}
where 
\begin{align*}
w_j (r, \bx, z) & = \frac{\pi_j p(\bx, z; \bfeta_j) p (r | \bx, z; \bxi_j)}{ \sum_{h = 1}^{H} \pi_h p(\bx, z; \bfeta_h) p(r | \bx, z; \bxi_h)}, \\
v_m (\bx, z) & = \frac{\pi_m p(\bx, z; \bfeta_m)}{\sum_{h = 1}^{H} \pi_h p(\bx, z; \bfeta_h)}. 
\end{align*}

The MC steps to compute the above integral are as follows:
\begin{enumerate}
\item Draw $l$ from a multinomial $\{1, \ldots, H \}$ with probabilities $( \pi_1, \ldots, \pi_H)$. 

\item Draw $\bx^{l}$ from $p(\bx; \bfeta_{l})$.

\item Draw $m$ from a multinomial $\{1, \ldots, H \}$ with probabilities $\{v_{1} (\bx^{l}, z), \ldots, v_{H} (\bx^{l}, z) \}$.

\item Draw $r^{m}$ from $p(r | \bx^{l}, z; \bxi_{m})$. 

\item If $r^{m} < K + 1$, calculate $ \sum_{h = 1}^{H} w_h (r^{m}, \bx^{l}, z) \{ \alpha^{(zm, h)} + \bx^{l} \bbeta^{(h)} \}$

\item If $r^{m} = K + 1$, calculate $E[Y | Z = z, R = k, \bX = \bx^{l}] =  \sum_{h = 1}^{H} w_h (k, \bx^{l}, z) \{ \alpha^{(zk, h)} + \bx^{l} \bbeta^{(h)} \}$ for $k = 1, \ldots, K$, compute $\alpha^z_{min}$, and then sample from the priors specified in Section 2.4.

\end{enumerate}

To assess goodness of fit in Section 3.3, we need to compute $E[Y | Z=z, R=r]$ as follows:
\begin{align*}
E[Y | Z = z, R = r] & = \int  E[Y | Z = z, R = r, \bX = \bx] dF(\bx | z, r) \\
& = \int \sum_{j = 1}^{H} w_j (r, \bx, z) ( \alpha^{(zr, j)} + \bx \bbeta^{(j)}) p(\bx | z, r) d\bx \\
& = \int \sum_{j = 1}^{H} w_j (r, \bx, z) ( \alpha^{(zr, j)} + \bx \bbeta^{(j)}) \sum_{h = 1}^{H} u_h (z, r) p(\bx; \bfeta_h) d\bx, 
\end{align*}
where
\begin{align*}
u_h (z, r) & = \frac{\pi_h p(z, r; \bfeta_{zh}, \bxi_h)}{\sum_{l = 1}^{H} \pi_l p(z, r; \bfeta_{zl}, \bxi_l)}\\
& = \frac{\pi_h p(z; \bfeta_{zh}, \bxi_{h}) p(r; \bxi_{rh})}{\sum_{l = 1}^{H} \pi_l p(z; \bfeta_{zl}) p(r; \bxi_{l})}
\end{align*}

 The MC steps to compute the above integral are as follows: 
\begin{enumerate}
\item Draw $l$ from a multinomial $\{1, \ldots, H \}$ with probabilities $\{u_1 (z, r), \ldots, u_{H} (z, r) \}$.

\item Draw $\bx^{l}$ from $p(\bx |z, r; \bfeta_{l}, \bxi_l)$.

\item Calculate $\sum_{h = 1}^{H} w_h (r, \bx^{l}, z) ( \alpha^{(zr, h)} + \bx^{l} \bbeta^{(h)})$.

\end{enumerate}

\section{Default Priors}
We adopt the specification of the hyperparameters of the distributions $H_{0 \cdot}$ from
\cite{linero2015} and \cite{Roy:2018} . The data $Y$ and $X_2$ is standardized with  mean 0 and  variance 0.5. For each mixture component $h$, we have
\begin{align*}
Y_i | R_i,  \bX_i,  Z_i; \bbeta_i^{(h)}, \alpha_i^{(zr, h)}, \sigma_i^{(2, h)} & \sim N(\alpha_i^{(zr, h)} +  \bX_i \bbeta_i^{(h)},\sigma_i^{(2, h)}), \\
R_i |  \bX_i, Z_i; \bxi_i^{(h)}  & \sim \mbox{Multinomial} (1,  \bxi_i^{(h)}), \\
X_{2i} | m_i,  \tau_i^{(h)} & \sim N(m_i^{(h)}, \tau_i^{(2, h)}), \\
\bX_{1i} | Z_i; \bfeta_i^{(h)} & \sim \mbox{Multinomial} (1,  \bfeta_i^{(h)}), \\
Z_i & \sim \mbox{Ber}(\eta_{zi}^{(h)}). 
\end{align*}
We specify $\alpha_i^{(zr, h)} \sim N(\mu_{\alpha_{z_i r_i}}, \sigma_{\alpha_{z_i}}^{2})$, $\bbeta_i^{(h)} \sim N(\mu_{\bbeta_{i}}, c \; \sigma_{\bbeta}^2)$, and $m_i \sim N(\mu_{m}, \sigma_{m}^2)$. We assume $\mu_{\alpha_{z_i r_i}}$, and $\mu_{m}$ follow a normal distribution with mean 0 and the variance 0.5. THe hyperparameters  $\mu_{\bbeta_{i}}$ and $\sigma_{\bbeta}^2$ are obtained from the estimated linear regression coefficients and the estimated variance of the coefficients. Here $c = \lceil 344 / 5 \rceil $, where 344 is the degrees of freedom. For the hyperparameters of the multinomial distributions, we set $\bxi_i^{(h)} \sim \text{DIR} (\mathbbm{1}_{K + 1} / (K + 1))$ and $\bfeta_i^{(h)} \sim \text{DIR} ( \mathbbm{1}_{4} / 4 ) $, where $\mathbbm{1}_{n} = (1, \ldots, 1) \in \mathbbm{R}^{n}$. In addition, $\eta_{zi}^{(h)} \sim \text{Beta} (1, 1)$. 
 
 The prior for the variance components were specified as follows.
\begin{align*}
\sigma_i^{(2, h)} & \sim   \Gamma^{-1} (S_1, S_1 W_1), & \qquad  S_1 - 2 & \sim  \Gamma^{-1}(1, 1), \\
W_1 & \sim  \Gamma(1, \frac{2}{G_Y}), & \qquad  \tau_i^{(2, h)} & \sim \Gamma^{-1} (S_2, S_2 W_2), \\
S_2 - 2 & \sim  \Gamma^{-1}(1, 1), & \qquad 
W_2  & \sim \Gamma(1, \frac{2}{G_{X_2}}), \\
\sigma_{\alpha_{z_i}}^{2} & \sim  \Gamma^{-1} (s_{z}, s_{z} \lambda_{z} g_{z}), & \qquad 
\lambda_{z}  & \sim  \Gamma(1, 1), \\
s_{z} - 2 & \sim  \Gamma^{-1}(1, 1), & \qquad 
\sigma_{m}^2 & \sim  \Gamma^{-1} (s_{2}, s_{2} \lambda_{2} G_{X_2}), \\
\lambda_{2}  & \sim  \Gamma(1, 1),  & \qquad 
s_{2} - 2 & \sim  \Gamma^{-1}(1, 1), \\  
\end{align*}
where $G_Y$ and $G_{X_2}$ are the variances of $Y$ and the baseline outcome, and $g_{z}$ is the conditional variances of $Y | Z = z$.  %{\bf Minji: this still needs clarifications.  are these conditional VARIANCES?  AND WHAT IS $G_1$ VS. $g_1$? }

\pagebreak

\section{Goodness of Fit}

\begin{table}[htpt]
\centering
\renewcommand{\arraystretch}{1.5}
\footnotesize
\caption{Goodness of fit: Expectations of the response given each merged attempt $R^\star$. 'Full' and 'Merged' correspond to the two conditional specifications for $Y$ in the DPM.}
\label{ch3:datatab2}

\begin{tabular}{@{}ccccccc@{}}
\hline
 & & & \multicolumn{2}{c}{Full} & \multicolumn{2}{c}{Merged}\\
\hline
$R^\star$ & Z &  $\bar{Y}$ & mean   &   95\% CI  & mean   &   95\% CI   \\
\hline

1 & 0 & 42.4 & 41.3 & (39.0, 43.4) & 41.5 & (39.3, 43.6) \\
 & 1 & 40.7 & 41.1 & (38.8, 43.3) & 41.2 & (38.9, 43.5) \\

\hline

2 & 0  & 41.3 & 40.9 & (39.0, 42.9)  & 40.9 & (39.0, 42.9) \\
 & 1 & 40.2 & 40.3 & (38.3, 42.3)  & 40.3 & (38.3, 42.2) \\
 
\hline

3 & 0 & 38.7 & 40.4 & (37.4, 43.0) & 40.1 & (37.1, 42.6) \\
& 1 & 38.6 & 40.2 & (37.3, 43.0) & 39.7 & (36.7, 42.3) \\

\hline

\end{tabular}
\end{table}

\pagebreak

\section{Additional Simulations}
\begin{table}[htpt]
\centering
\renewcommand{\arraystretch}{1.5}
\footnotesize
\caption{Scenario 1 with normal distribution: Bias, MSE and interval coverage probability and length for the estimated treatment effect $\theta$, based on 1000 replications. The sample size is 409 and 1000, respectively. The subscript 4 represents that subjects with 3 to 9 attempts were merged to 3 attempts ($K = 3$). Prior equal to 'none' corresponds to the estimate of the treatment effect without the missing group $K+1$ (so completers only). The notation for the different priors can be found in Section 3.2.  }
\label{scen1:DPM:normal}

\begin{tabular}{@{}ccccccccccccc@{}}

\hline
Scenario &  \multicolumn{3}{c}{ 409$_4$ } & \multicolumn{3}{c}{ 409 }   &  \multicolumn{3}{c}{ 1000$_4$ } & \multicolumn{3}{c}{ 1000 } \\
%1000
Prior & bias  &  MSE &  coverage & bias  &  MSE  &  coverage & bias  &  MSE  &  coverage & bias  &  MSE  &  coverage\\
\hline

None & 0.034 & 1.145 & 0.951 & 0.033 & 1.163 & 0.954 & -0.012 & 0.451 & 0.952 & -0.012 & 0.452 & 0.958 \\

p.m & -0.053 & 1.159 & 0.953 & -0.031 && 1.167  0.954& -0.091  &0.458 & 0.959 & -0.056 & 0.458 & 0.957 \\

unif$_{10}$  & -0.052 & 1.159 & 0.953 & -0.030  &1.167 & 0.955 & -0.090 & 0.458 & 0.960 & -0.055 & 0.458 & 0.957\\

unif$_{20}$  & -0.051 & 1.159 & 0.953 & -0.029 & 1.167 & 0.955 & -0.089 & 0.458 &  0.960 & -0.053 & 0.458 & 0.959 \\

tri1$_{10}$  & -0.051 & 1.159 & 0.953 & -0.029 & 1.167 & 0.955 & -0.089 & 0.458  &0.960 & -0.054 & 0.458 & 0.959 \\

tri1$_{20}$  & -0.050 & 1.159 & 0.953 & -0.028 & 1.167 & 0.955 & -0.088 & 0.458 & 0.960 & -0.052 & 0.458 & 0.959 \\

tri2$_{10}$  & -0.052 & 1.159  & 0.953  & -0.030 & 1.167 & 0.955 & -0.090 & 0.458 & 0.960 & -0.055 & 0.458 & 0.959 \\

tri2$_{20}$  & -0.051 & 1.159 & 0.954 & -0.029 & 1.167 & 0.955 & -0.089 & 0.458 & 0.960 & -0.054 & 0.458 & 0.958 \\

\hline

\end{tabular}
\end{table}

\begin{table}[htpt]
\centering
\renewcommand{\arraystretch}{1.5}
\footnotesize
\caption{Scenario 1 with student's t distribution with 3 degrees of freedom : Bias, MSE and interval coverage probability and length for the estimated treatment effect $\theta$, based on 1000 replications. The sample size is 409 and 1000, respectively. The subscript 4 represents that subjects with 3 to 9 attempts were merged to 3 attempts ($K = 3$). Prior equal to 'none' corresponds to the estimate of the treatment effect without the missing group $K+1$ (so completers only). The notation for the different priors can be found in Section 3.2. }
\label{sup:scen1:DPM:tdist}
\begin{tabular}{@{}ccccccccccccc@{}}

\hline
Scenario &  \multicolumn{3}{c}{ 409$_4$ } & \multicolumn{3}{c}{ 409 }   &  \multicolumn{3}{c}{ 1000$_4$ } & \multicolumn{3}{c}{ 1000 } \\

Prior & bias  &  MSE &  coverage & bias  &  MSE  &  coverage & bias  &  MSE  &  coverage & bias  &  MSE  &  coverage\\
\hline

None & -0.017 & 0.699 & 0.944  &  0.015 & 0.734 & 0.951 & 0.036 &  0.262 & 0.960 & 0.087 & 0.273 & 0.967 \\

p.m & -0.100 & 0.713 & 0.952 & -0.046 & 0.744 & 0.953 & -0.044 & 0.264 & 0.967 & 0.046  & 0.271&  0.969  \\

unif$_{10}$  & -0.099 & 0.713 & 0.951 & -0.045 & 0.745 & 0.953 & -0.043 & 0.264 & 0.967 & 0.048 & 0.271 & 0.968 \\

unif$_{20}$  & -0.099 & 0.713 & 0.951 & -0.043 & 0.745 & 0.953 & -0.041 & 0.264 & 0.968 & 0.046 & 0.272 & 0.968  \\

tri1$_{10}$  & -0.099 & 0.713 & 0.951 & -0.044 & 0.745 & 0.953 & -0.042 & 0.264 & 0.967 & 0.048 & 0.272 & 0.968 \\

tri1$_{20}$  & -0.098 & 0.713 & 0.953 & -0.043 & 0.746 & 0.954 & -0.040&  0.264 & 0.968 & 0.051 & 0.272 & 0.969  \\

tri2$_{10}$  & -0.100 & 0.713 & 0.951 & -0.045 & 0.744 & 0.953 & -0.043 & 0.264 & 0.967 & 0.047 & 0.271 & 0.968  \\

tri2$_{20}$  & -0.099 & 0.713 & 0.951 & -0.044  & 0.745 & 0.954 & -0.042  & 0.264 & 0.968 & 0.048 & 0.272 & 0.968   \\

\hline

\end{tabular}
\end{table}

\begin{table}[htpt]
\centering
\renewcommand{\arraystretch}{1.5}
\footnotesize
\caption{Scenario 1 with skewed normal distribution : Bias, MSE and interval coverage probability and length for the estimated treatment effect $\theta$, based on 1000 replications. The sample size is 409 and 1000, respectively. The subscript 4 represents that subjects with 3 to 9 attempts were merged to 3 attempts ($K = 3$). Prior equal to 'none' corresponds to the estimate of the treatment effect without the missing group $K+1$ (so completers only). The notation for the different priors can be found in Section 3.2. }
\label{sup:scen1:DPM:skewed}

\begin{tabular}{@{}ccccccccccccc@{}}
\hline
Scenario &  \multicolumn{3}{c}{ 409$_4$ } & \multicolumn{3}{c}{ 409 }   &  \multicolumn{3}{c}{ 1000$_4$ } & \multicolumn{3}{c}{ 1000 } \\

Prior & bias  &  MSE &  coverage & bias  &  MSE  &  coverage & bias  &  MSE  &  coverage & bias  &  MSE  &  coverage\\
\hline

None & -0.047 & 0.506 & 0.947  &  -0.042 & 0.509 & 0.957 & -0.030 & 0.209 & 0.945 & -0.021 & 0.208 & 0.966 \\

p.m & -0.128 & 0.521 & 0.949 & -0.086 & 0.515 & 0.958 & -0.106 & 0.218 & 0.957 & -0.048 & 0.211 & 0.964 \\

unif$_{10}$  & -0.126 & 0.521 & 0.950 & -0.085 & 0.515 & 0.959 & -0.104 & 0.218 & 0.957 & -0.046 & 0.211 & 0.964\\

unif$_{20}$  &-0.125 & 0.520 & 0.949 & -0.083 & 0.515 & 0.960 & -0.102 & 0.218 & 0.957 & -0.044 & 0.211 & 0.965 \\

tri1$_{10}$  & -0.126 & 0.520 & 0.950 & -0.084 & 0.515  &0.960 & -0.103 & 0.218 & 0.957 & -0.046 & 0.211 & 0.965 \\

tri1$_{20}$  & -0.124 & 0.520 & 0.948 & -0.082 & 0.515 & 0.960 & -0.101 & 0.218 & 0.958 & -0.043 & 0.211 & 0.965 \\

tri2$_{10}$  & -0.127 & 0.521 & 0.950 & -0.085 & 0.515 & 0.959 & -0.104 & 0.218 & 0.957 & -0.047 & 0.211 & 0.964 \\

tri2$_{20}$  & -0.126 & 0.520 & 0.950 & -0.084 & 0.515 & 0.960 & -0.103 & 0.218 & 0.957 & -0.046 & 0.211&  0.965 \\

\hline

\end{tabular}
\end{table}

\begin{table}[htpt]
\centering
\renewcommand{\arraystretch}{1.5}
\footnotesize
\caption{Scenario 1 with PMM-RAM : Bias, MSE and coverage probability (and interval length) for the estimated treatment effect $\theta$, based on 1000 replications. The sample size is 409 and 1000, respectively. The sensitivity parameter is set to $C = 3$ for the RAM-PMM.}
\label{sup:scen1:PMM:collapsed}

\begin{tabular}{@{}ccccccc@{}}

\hline
Scenario &  \multicolumn{3}{c}{ 409 } &  \multicolumn{3}{c}{ 1000 } \\

 & bias &  MSE &  coverage & bias &  MSE  &  coverage \\
\hline

Normal & 0.028 & 1.313 & 0.954  & -0.032 & 0.507 & 0.956 \\

student's t & -0.007 & 1.270 & 0.956 & 0.038 & 0.558 & 0.961 \\

skewed & -0.032 & 0.571 & 0.949 & -0.036 & 0.233 & 0.963 \\

\hline

\end{tabular}
\end{table}

\begin{table}[htpt]
\centering
\renewcommand{\arraystretch}{1.5}
\footnotesize
\caption{Scenario 1 with skewed normal distribution for PMM-RAM and SM; Bias, MSE and coverage probability for the estimated treatment effect $\theta$, based on 1000 samples. The sample size is 409 and 1000, respectively. The sensitivity parameter is set to $C = 3$.}
\label{sup:scen1:PMM:SM}

\begin{tabular}{@{}ccccccc@{}}

\hline
Scenario &  \multicolumn{3}{c}{ 409 } &  \multicolumn{3}{c}{ 1000 } \\

 & bias &  MSE &  coverage & bias &  MSE  &  coverage \\
\hline

PMM & -0.160 & 0.547 & 0.954 & -0.153 & 0.227 & 0.961 \\

SM & -0.163 & 0.572 & 0.927 & -0.160 & 0.244 & 0.934 \\

\hline

\end{tabular}
\end{table}

\begin{table}[htpt]
\centering
\renewcommand{\arraystretch}{1.5}
\footnotesize
\caption{Scenario 2 : Bias, MSE and interval coverage probability and length for the estimated treatment effect $\theta$, based on 1000 replications. The sample size is 500 and 1000, respectively. The subscript 4 represents that subjects with 3 to 9 attempts were merged to 3 attempts ($K = 3$). Prior equal to 'none' corresponds to the estimate of the treatment effect without the missing group $K+1$ (so completers only). The notation for the different priors can be found in Section 3.2. }
\label{sup:scen2:DPM:normal}

\begin{tabular}{@{}ccccccccccccc@{}}

\hline
Scenario &  \multicolumn{3}{c}{ 500$_4$ } & \multicolumn{3}{c}{ 500 }   &  \multicolumn{3}{c}{ 1000$_4$ } & \multicolumn{3}{c}{ 1000 } \\

Prior & bias  &  MSE &  coverage & bias  &  MSE  &  coverage & bias  &  MSE  &  coverage & bias  &  MSE  &  coverage\\
\hline

None & 0.017 & 0.832 & 0.947 & 0.019 & 0.839 & 0.951 & -0.015 & 0.409 & 0.957 & -0.013 & 0.411 & 0.956 \\

p.m & 0.026 & 0.828 & 0.951 & 0.024 & 0.841 & 0.957 & -0.014 & 0.413 & 0.956 & -0.020 & 0.418 & 0.960 \\

unif$_{10}$  & 0.026 & 0.828 & 0.952 & 0.023 & 0.841&  0.957 & -0.015 & 0.413 & 0.956 & -0.022 & 0.419 & 0.961 \\

unif$_{20}$  & 0.025 & 0.829 & 0.952 & 0.022 & 0.842 & 0.958 & -0.017 & 0.414 & 0.956 & -0.024 & 0.420 & 0.961 \\

tri1$_{10}$  & 0.025 & 0.828 & 0.952 & 0.023 & 0.841&  0.957 & -0.016 & 0.413 & 0.956 & -0.023 & 0.419 & 0.961 \\

tri1$_{20}$  & 0.024 & 0.829 & 0.952 & 0.022 & 0.842 & 0.958 & -0.018 & 0.414 & 0.956 & -0.025 & 0.420 & 0.961 \\

tri2$_{10}$  & 0.026&  0.828  & 0.952 & 0.024  & 0.841&  0.957 & -0.015 & 0.413 & 0.956 & -0.021 & 0.418 & 0.960 \\

tri2$_{20}$  & 0.025 & 0.828 & 0.952 & 0.023 & 0.841 & 0.958 & -0.016 & 0.413 & 0.956 & -0.023 & 0.419 & 0.961 \\

\hline

\end{tabular}
\end{table}

\begin{table}[htpt]
\centering
\renewcommand{\arraystretch}{1.5}
\footnotesize
\caption{Scenario 2 with student's t distribution with 3 degrees of freedom : Bias, MSE and interval coverage probability and length for the estimated treatment effect $\theta$, based on 1000 replications. The sample size is 500 and 1000, respectively. The subscript 4 represents that subjects with 3 to 9 attempts were merged to 3 attempts ($K = 3$). Prior equal to 'none' corresponds to the estimate of the treatment effect without the missing group $K+1$ (so completers only). The notation for the different priors can be found in Section 3.2. }
\label{sup:scen2:DPM:tdist}
\begin{tabular}{@{}ccccccccccccc@{}}

\hline
Scenario &  \multicolumn{3}{c}{ 500$_4$ } & \multicolumn{3}{c}{ 500 }   &  \multicolumn{3}{c}{ 1000$_4$ } & \multicolumn{3}{c}{ 1000 } \\

Prior & bias  &  MSE &  coverage & bias  &  MSE  &  coverage & bias  &  MSE  &  coverage & bias  &  MSE  &  coverage\\
\hline

None & 0.012 & 0.471 & 0.956  &  0.007 & 0.605 & 0.957 & 0.036 & 0.243 & 0.956 & 0.025 & 0.247 & 0.959 \\

p.m & 0.020 & 0.473 & 0.958 & 0.010 & 0.499 & 0.959 & 0.039 & 0.246 & 0.959 &0.020 & 0.252 & 0.961 \\

unif$_{10}$  & 0.019 & 0.473 & 0.957 & 0.009  0.500  0.959 & 0.038 & 0.246 & 0.959 &0.019 & 0.252 & 0.962 \\

unif$_{20}$  & 0.018 & 0.474 & 0.956 & 0.008  0.501  0.959 & 0.036  & 0.246 & 0.959 & 0.017 & 0.253 & 0.962 \\

tri1$_{10}$  & 0.019 & 0.474 & 0.957 & 0.009 & 0.500 & 0.959 & 0.037 & 0.246 & 0.959 & 0.018  & 0.253 & 0.962 \\

tri1$_{20}$  & 0.017 & 0.474 & 0.956 & 0.007 & 0.502 & 0.959 & 0.035 & 0.246 & 0.960 & 0.016 & 0.253 & 0.962 \\

tri2$_{10}$  & 0.019 & 0.473 & 0.957 & 0.010 & 0.500 & 0.959 & 0.038 & 0.246 & 0.959 & 0.019 & 0.252 & 0.962\\

tri2$_{20}$  & 0.019 & 0.474 & 0.957 & 0.009 & 0.500 & 0.959 & 0.037 & 0.246 & 0.959 & 0.018 &  0.253 & 0.962  \\

\hline

\end{tabular}
\end{table}

\begin{table}[htpt]
\centering
\renewcommand{\arraystretch}{1.5}
\footnotesize
\caption{Scenario 2 with PMM-RAM : Bias, MSE and coverage probability (and interval length) for the estimated treatment effect $\theta$, based on 1000 replications. The sample size is 500 and 1000, respectively. The sensitivity parameter is set to $C = 3$ for the RAM-PMM.}
\label{sup:scen2:PMM:collapsed}
%collapsed
\begin{tabular}{@{}ccccccc@{}}

\hline
Scenario &  \multicolumn{3}{c}{ 500 } &  \multicolumn{3}{c}{ 1000 } \\

 & bias &  MSE &  coverage & bias &  MSE  &  coverage \\
\hline

Normal & -0.044 & 0.282 & 0.947  & -0.060 & 0.136 & 0.945 \\

student's t & -0.054 & 1.098 & 0.960 & -0.006 & 0.566 & 0.952 \\

skewed & -0.062 & 0.479 & 0.953 & -0.079 & 0.247 & 0.942 \\

\hline

\end{tabular}
\end{table}

\begin{table}[htpt]
\centering
\renewcommand{\arraystretch}{1.5}
\footnotesize
\caption{Scenario 3: Bias, MSE and interval coverage probability and length for the estimated treatment effect $\theta$, based on 1000 replications. The sample size is 500 and 1000, respectively. The subscript 4 represents that subjects with 3 to 9 attempts were merged to 3 attempts ($K = 3$). Prior equal to 'none' corresponds to the estimate of the treatment effect without the missing group $K+1$ (so completers only). The notation for the different priors can be found in Section 3.2. }
\label{sup:scen3:DPM:normal}

\begin{tabular}{@{}ccccccccccccc@{}}

\hline
Scenario &  \multicolumn{3}{c}{ 500$_4$ } & \multicolumn{3}{c}{ 500 }   &  \multicolumn{3}{c}{ 1000$_4$ } & \multicolumn{3}{c}{ 1000 } \\

Prior & bias  &  MSE &  coverage & bias  &  MSE  &  coverage & bias  &  MSE  &  coverage & bias  &  MSE  &  coverage\\
\hline

None & 0.017 & 0.938 & 0.962 & 0.013 & 0.911 & 0.963 & -0.012 & 0.475 & 0.966 & -0.012 & 0.445 & 0.971 \\

p.m & 0.027 & 0.951 & 0.962 & 0.028 & 0.923 & 0.971 & -0.009 & 0.487 & 0.964 & -0.003 & 0.464 & 0.973 \\

unif$_{10}$  & 0.029 & 0.953 & 0.964 & 0.030 & 0.927 & 0.971 & -0.007 & 0.489 & 0.964 & -0.002 &  0.466 & 0.974 \\

unif$_{20}$  & 0.031 & 0.955 & 0.964 & 0.032 & 0.930 & 0.971 & -0.006 & 0.490 & 0.965 & 0.000 &  0.468  &0.973 \\

tri1$_{10}$  & 0.030 & 0.954 & 0.964 & 0.031 & 0.928  &0.971 & -0.007 & 0.489 & 0.965 & -0.001 & 0.466 & 0.974 \\

tri1$_{20}$  & 0.032 & 0.956 & 0.964 & 0.034 & 0.933 & 0.971 & -0.005 & 0.491 & 0.966 & 0.001 & 0.469 & 0.976 \\

tri2$_{10}$  & 0.029 & 0.952 & 0.963 & 0.029 & 0.925 & 0.971 & -0.008 & 0.488 & 0.964 & -0.002 & 0.465 & 0.974 \\

tri2$_{20}$  & 0.030 & 0.954 & 0.964 & 0.031 & 0.928 & 0.971 & -0.007 & 0.489 & 0.965 & -0.001 & 0.466 & 0.974 \\

\hline

\end{tabular}
\end{table}

\begin{table}[htpt]
\centering
\renewcommand{\arraystretch}{1.5}
\footnotesize
\caption{Scenario 3 with student's t distribution with 3 degrees of freedom: Bias, MSE and interval coverage probability and length for the estimated treatment effect $\theta$, based on 1000 replications. The sample size is 500 and 1000, respectively. The subscript 4 represents that subjects with 3 to 9 attempts were merged to 3 attempts ($K = 3$). Prior equal to 'none' corresponds to the estimate of the treatment effect without the missing group $K+1$ (so completers only). The notation for the different priors can be found in Section 3.2. }
\label{sup:scen3:DPM:tdist}

\begin{tabular}{@{}ccccccccccccc@{}}

\hline
Scenario &  \multicolumn{3}{c}{ 500$_4$ } & \multicolumn{3}{c}{ 500 }   &  \multicolumn{3}{c}{ 1000$_4$ } & \multicolumn{3}{c}{ 1000 } \\

Prior & bias  &  MSE &  coverage & bias  &  MSE  &  coverage & bias  &  MSE  &  coverage & bias  &  MSE  &  coverage\\
\hline

None & 0.011 & 0.628 & 0.971 &  0.017 & 0.610 & 0.975 & 0.047 & 0.352 & 0.969 & 0.037 & 0.335 & 0.979 \\

p.m & 0.027 & 0.647 & 0.974 & 0.029 & 0.635 & 0.977 & 0.049 & 0.360 & 0.971 & 0.038 & 0.327 & 0.982 \\

unif$_{10}$  & 0.028 & 0.649 & 0.975 & 0.031 & 0.639 & 0.978 & 0.050 & 0.361 & 0.972 & 0.040 & 0.328 & 0.982\\

unif$_{20}$  & 0.030 & 0.651 & 0.975 & 0.033 & 0.645&  0.979  & 0.052 & 0.361 & 0.972 & 0.041 & 0.330 & 0.982 \\

tri1$_{10}$  & 0.029 & 0.650 & 0.975 & 0.032 & 0.641 & 0.978 & 0.051 & 0.361 & 0.972 & 0.040 & 0.329 & 0.982 \\

tri1$_{20}$  & 0.031 & 0.652 & 0.975 & 0.034 & 0.648 & 0.979 & 0.053 & 0.363 & 0.972 & 0.042 & 0.331 & 0.982 \\

tri2$_{10}$  & 0.028 & 0.648 & 0.975 & 0.031 & 0.638 & 0.979 & 0.050 & 0.361 & 0.972 & 0.039 &  0.328 & 0.982 \\

tri2$_{20}$  & 0.029 & 0.650 & 0.975 & 0.032 & 0.641 & 0.978 & 0.051 & 0.361 & 0.972 & 0.040 & 0.329 & 0.982  \\

\hline

\end{tabular}
\end{table}

\begin{table}[htpt]
\centering
\renewcommand{\arraystretch}{1.5}
\footnotesize
\caption{Scenario 3 with skewed normal distribution : Bias, MSE and interval coverage probability and length for the estimated treatment effect $\theta$, based on 1000 replications. The sample size is 500 and 1000, respectively. The subscript 4 represents that subjects with 3 to 9 attempts were merged to 3 attempts ($K = 3$). Prior equal to 'none' corresponds to the estimate of the treatment effect without the missing group $K+1$ (so completers only). The notation for the different priors can be found in Section 3.2. }
\label{sup:scen3:DPM:skewed}

\begin{tabular}{@{}ccccccccccccc@{}}

\hline
Scenario &  \multicolumn{3}{c}{ 500$_4$ } & \multicolumn{3}{c}{ 500 }   &  \multicolumn{3}{c}{ 1000$_4$ } & \multicolumn{3}{c}{ 1000 } \\

Prior & bias  &  MSE &  coverage & bias  &  MSE  &  coverage & bias  &  MSE  &  coverage & bias  &  MSE  &  coverage\\
\hline

None & -0.015 & 0.437 & 0.959 & 0.020 & 0.924 & 0.961 & -0.029 & 0.248 & 0.974 & -0.012 & 0.444 & 0.969 \\

p.m & -0.007 & 0.448 & 0.962 & 0.035 & 0.936 & 0.970 & -0.025 & 0.247 & 0.970 & -0.003 & 0.463  &0.974 \\

unif$_{10}$  & -0.005 & 0.449 & 0.963 & 0.037 & 0.939 & 0.970 & -0.023 & 0.247 & 0.971 &  -0.001 & 0.465 & 0.975 \\

unif$_{20}$  & -0.004 & 0.450 & 0.964 & 0.039 & 0.943 & 0.973 & -0.022 & 0.248 & 0.972 &  0.001  & 0.467 & 0.975 \\

tri1$_{10}$  & -0.005 & 0.449 & 0.963 & 0.038 & 0.940 & 0.970 & -0.023 & 0.247 & 0.972 &  -0.001 & 0.465 & 0.975 \\

tri1$_{20}$  & -0.003 & 0.451 & 0.963 & 0.041 & 0.945 & 0.973 & -0.021 & 0.248 & 0.972 &  0.002 & 0.468 & 0.975 \\

tri2$_{10}$  & -0.006 & 0.449 & 0.962 & 0.037 & 0.938 & 0.970 & -0.024 & 0.247 & 0.972 &  -0.002 & 0.464 & 0.975 \\

tri2$_{20}$  & -0.005 & 0.450 & 0.963 & 0.038 & 0.940 & 0.970 & -0.023 & 0.247 & 0.971 & 0.000 & 0.465 & 0.975 \\

\hline

\end{tabular}
\end{table}

\begin{table}[htpt]
\centering
\renewcommand{\arraystretch}{1.5}
\footnotesize
\caption{Scenario 3 with PMM-RAM : Bias, MSE and coverage probability (and interval length) for the estimated treatment effect $\theta$, based on 1000 replications. The sample size is 500 and 1000, respectively. The sensitivity parameter is set to $C = 3$ for the RAM-PMM.}
\label{sup:scen3:PMM}

\begin{tabular}{@{}ccccccc@{}}

\hline
Scenario &  \multicolumn{3}{c}{ 500 } &  \multicolumn{3}{c}{ 1000 } \\

 &  bias &  MSE &  coverage & bias &  MSE  &  coverage \\
\hline

Normal & 0.062 & 1.622 & 0.948  & 0.046 & 0.782 & 0.957 \\

student's t & 0.041 & 1.659 & 0.952 & 0.118 & 0.812 & 0.949 \\

skewed & 0.033 & 1.038 & 0.949 & 0.041 & 0.509 & 0.942 \\

\hline

\end{tabular}
\end{table}

\begin{table}[htpt]
\centering
\renewcommand{\arraystretch}{1.5}
\footnotesize
\caption{Scenario 3 with skewed normal distribution for PMM-RAM and selection model (SM) : Bias, MSE and coverage probability (and interval length) for the estimated treatment effect $\theta$, based on 1000 replications. The sample size is 500 and 1000, respectively. The sensitivity parameter is set to $C = 3$ for the RAM-PMM.}
\label{sup:scen3:PMM:SM}

\begin{tabular}{@{}ccccccc@{}}

\hline
Scenario &  \multicolumn{3}{c}{ 500 } &  \multicolumn{3}{c}{ 1000 } \\

 & bias &  MSE &  coverage & bias &  MSE  &  coverage \\
\hline

PMM & -0.019 & 0.613 & 0.959 & -0.014 & 0.296 & 0.945 \\

SM & -0.027 & 0.789 & 0.953 & -0.020 & 0.385 & 0.945 \\

\hline

\end{tabular}
\end{table}

\begin{table}[htpt]
\centering
\renewcommand{\arraystretch}{1.5}
\footnotesize
\caption{Scenario 4: Bias, MSE and interval coverage probability and length for the estimated treatment effect $\theta$, based on 1000 replications. The sample size is 500 and 1000, respectively. The subscript 4 represents that subjects with 3 to 9 attempts were merged to 3 attempts ($K = 3$). Prior equal to 'none' corresponds to the estimate of the treatment effect without the missing group $K+1$ (so completers only). The notation for the different priors can be found in Section 3.2. }
\label{scen4:DPM}

\begin{tabular}{@{}ccccccccccccc@{}}

\hline
Scenario &  \multicolumn{3}{c}{ 500$_4$ } & \multicolumn{3}{c}{ 1000$_4$ }   &  \multicolumn{3}{c}{ 500 } & \multicolumn{3}{c}{ 1000 } \\

Prior & bias  &  MSE &  coverage & bias  &  MSE  &  coverage & bias  &  MSE  &  coverage & bias  &  MSE  &  coverage\\
\hline

None & -0.167 & 0.890 & 0.941 & -0.158 & 0.428  &  0.962 & -0.140 & 0.885 & 0.946 & -0.145  &  0.422 &  0.957 \\

p.m & -0.165 & 0.885 & 0.940 & -0.152 & 0.427 & 0.955 & -0.146&  0.885 & 0.949 & -0.142 & 0.424 & 0.964 \\

unif$_{10}$  & -0.165 & 0.886 & 0.940 & -0.152 & 0.427 & 0.955 & -0.146 & 0.885 & 0.948 & -0.142 & 0.425 & 0.965 \\

unif$_{20}$  & -0.165 & 0.886 & 0.940 &-0.152 & 0.427 & 0.955 & -0.147 & 0.885 & 0.948 & -0.143 & 0.425 & 0.965\\

tri1$_{10}$  & -0.165 & 0.886 & 0.940 & -0.152 & 0.427 & 0.955 & -0.147 & 0.885 & 0.948 & -0.143 & 0.425 & 0.965\\

tri1$_{20}$  & -0.165 & 0.886 & 0.940 & -0.152 & 0.427 & 0.954  & -0.147 & 0.885 & 0.949 & -0.144 & 0.425 & 0.965 \\

tri2$_{10}$  & -0.165 & 0.885 & 0.940 & -0.152 & 0.427 & 0.955 &-0.146 & 0.885&  0.949 & -0.142 & 0.425 & 0.964 \\

tri2$_{20}$  & -0.165 & 0.886 & 0.940 & -0.152 & 0.427 & 0.955 & -0.147 & 0.885 & 0.948 &-0.143   & 0.425 & 0.965 \\

\hline

\end{tabular}
\end{table}

\begin{table}[htpt]
\centering
\renewcommand{\arraystretch}{1.5}
\footnotesize
\caption{Scenario 4 with PMM-RAM; Bias, MSE and coverage probability for the estimated treatment effect $\theta$, based on 1000 samples. The sample size is 500 and 1000, respectively. The subscript 4 represents that subjects with 3 to 8 attempts were merged to 3 attempts. ($K = 3$). The sensitivity parameter is set to $C = 3$.%{\bf The result without collapsing attempts is in last two columns}
}
\label{sup:scen4:PMM}

\begin{tabular}{@{}ccccccccccccc@{}}

\hline
Scenario &  \multicolumn{3}{c}{ 500$_4$ } &  \multicolumn{3}{c}{ 1000$_4$ } &  \multicolumn{3}{c}{ 500 } &  \multicolumn{3}{c}{ 1000 } \\

 & bias  &  MSE &  coverage & bias  &  MSE  &  coverage & bias  &  MSE &  coverage & bias  &  MSE  &  coverage \\
\hline

 & -0.027   & 0.799 & 0.952  & -0.011   & 0.366 & 0.970 & -0.039   & 0.801 & 0.956 & -0.017   & 0.369 & 0.976 \\

\hline

\end{tabular}
\end{table}

\begin{table}[htpt]
\centering
\renewcommand{\arraystretch}{1.5}
\footnotesize
\caption{Scenario 4 with SM; Bias, MSE and coverage probability for the estimated treatment effect $\theta$, based on 1000 samples. The sample size is 500 and 1000, respectively. The subscript 4 represents that subjects with 3 to 8 attempts were merged to 3 attempts. ($K = 3$).}
\label{sup:scen4:SM}

\begin{tabular}{@{}ccccccccccccc@{}}

\hline
Scenario &  \multicolumn{3}{c}{ 500$_4$ } &  \multicolumn{3}{c}{ 1000$_4$ } &  \multicolumn{3}{c}{ 500 } &  \multicolumn{3}{c}{ 1000 } \\

 & bias  &  MSE &  coverage & bias  &  MSE  &  coverage & bias  &  MSE &  coverage & bias  &  MSE  &  coverage \\
\hline

 & -0.054   & 0.945 & 0.944  & -0.016   & 0.435 & 0.960 & -0.041   & 0.921 & 0.946 & -0.019   & 0.430 & 0.961 \\

\hline

\end{tabular}
\end{table}

\begin{table}[htpt]
\centering
\renewcommand{\arraystretch}{1.5}
\footnotesize
\caption{Scenario 4 : Bias, MSE and interval coverage probability and length for the estimated treatment effect $\theta$, based on 1000 replications. The sample size is 500, 1000, and 10000, respectively. Prior equal to 'none' corresponds to the estimate of the treatment effect without the missing group $K+1$ (so completers only). The notation for the different priors can be found in Section 3.2. }
\label{sup:scen4:DPM:large}

\begin{tabular}{@{}cccccccccc@{}}

\hline
Scenario &   \multicolumn{3}{c}{ 500 } & \multicolumn{3}{c}{ 1000 } & \multicolumn{3}{c}{ 10000 } \\

Prior &  bias  &  MSE  &  coverage & bias  &  MSE  &  coverage & bias  &  MSE  &  coverage\\
\hline

None &  -0.140 & 0.885 & 0.946 & -0.145  &  0.422 &  0.957  & -0.124  &   0.059 &    0.959 \\

p.m &  -0.146&  0.885 & 0.949 & -0.142 & 0.424 & 0.964 & -0.105  &   0.057 &    0.972\\

unif$_{10}$  & -0.146 & 0.885 & 0.948 & -0.142 & 0.425 & 0.965 & -0.105  &   0.057 &    0.972\\

unif$_{20}$  & -0.147 & 0.885 & 0.948 & -0.143 & 0.425 & 0.965 & -0.105  &   0.057 &    0.976\\

tri1$_{10}$  &  -0.147 & 0.885 & 0.948 & -0.143 & 0.425 & 0.965 & -0.105  &   0.057 &    0.974\\

tri1$_{20}$  & -0.147 & 0.885 & 0.949 & -0.144 & 0.425 & 0.965 & -0.105  &   0.057 &    0.978\\

tri2$_{10}$  &  -0.146 & 0.885&  0.949 & -0.142 & 0.425 & 0.964 & -0.105  &   0.057 &    0.972\\

tri2$_{20}$  & -0.147 & 0.885 & 0.948 &-0.143   & 0.425 & 0.965  & -0.105  &   0.057 &    0.973\\

\hline

\end{tabular}
\end{table}

%RR
\begin{table}[htpt]
\centering
\renewcommand{\arraystretch}{1.5}
\footnotesize
\caption{Scenario 4 without restriction on attempts : Bias, MSE and interval coverage probability and length for the estimated treatment effect $\theta$ based on 1000 replications. The sample size is 500 and 1000, respectively. The subscript 4 represents that subjects with 3 to 9 attempts were merged to 3 attempts ($K = 3$). Prior equal to 'none' corresponds to the estimate of the treatment effect without the missing group $K+1$ (so completers only). The notation for the different priors can be found in Section 3.2.}
\label{sup:scen4:DPM:sparse}
%1000
\begin{tabular}{@{}ccccccc@{}}

\hline
Scenario &   \multicolumn{3}{c}{ 500 } & \multicolumn{3}{c}{ 1000 } \\

Prior &  bias  &  MSE  &  coverage & bias  &  MSE  &  coverage \\
\hline

None &  -0.137 &  0.883 &  0.950  & -0.146 &  0.420 &  0.958  \\

p.m &  -0.141 & 0.889 & 0.954  & -0.144 & 0.420 & 0.962\\

unif$_{10}$  &-0.142 & 0.889 & 0.954  &  -0.145 & 0.420  & 0.962\\

unif$_{20}$  & -0.143 & 0.889 & 0.954  &  -0.146 & 0.420 & 0.964 \\

tri1$_{10}$  &  -0.142 & 0.889 & 0.954  &  -0.145 & 0.420 & 0.963\\

tri1$_{20}$  &  -0.143 & 0.890 & 0.954 & -0.146 & 0.420 & 0.964 \\

tri2$_{10}$  &-0.142 & 0.889 & 0.954  & -0.145 & 0.420 & 0.963\\

tri2$_{20}$  & -0.142 & 0.889 & 0.954  & -0.145 & 0.420  & 0.964 \\

\hline

\end{tabular}
\end{table}

\begin{table}[htpt]
\centering
\renewcommand{\arraystretch}{1.5}
\footnotesize
\caption{Scenario 4 without restriction on attempts for PMM-RAM : Bias, MSE and coverage probability (and interval length) for the estimated treatment effect $\theta$, based on 1000 replications. The sample size is 500 and 1000, respectively. The sensitivity parameter is set to $C = 3$ for the RAM-PMM.}
\label{sup:scen4:PMM:sparse}

\begin{tabular}{@{}ccccccc@{}}

\hline
Scenario &   \multicolumn{3}{c}{ 500 } &  \multicolumn{3}{c}{ 1000 } \\

 & bias  &  MSE &  coverage & bias  &  MSE  &  coverage  \\
\hline

 & -0.028   & 0.816 & 0.952  & -0.012   & 0.364 & 0.978 \\

\hline

\end{tabular}
\end{table}

\begin{table}[htpt]
\centering
\renewcommand{\arraystretch}{1.5}
\scriptsize
\caption{Scenario 6: Bias, MSE, coverage probability, interval length for the estimated treatment effect $\theta$, based on 1000 replications.  The 
  the sample size is 500.}
\label{sup:scen6:N=500}

\begin{tabular}{@{}cccccccccc@{}}
\hline missing \% & & none & point mass & unif$_{10}$ & unif$_{20}$ &   tri1$_{10}$ & tri1$_{20}$  &  tri2$_{10}$ & tri2$_{20}$    \\
\hline
      10 \% & bias & -0.018 & -0.017 & -0.018 & -0.019 & -0.018 & -0.019 & -0.017 & -0.018 \\
      & MSE & 0.896 & 0.900 & 0.901 & 0.902 & 0.901 & 0.902  & 0.901  & 0.901\\
      & coverage & 0.960& 0.954 & 0.955 & 0.954 & 0.953 & 0.955 & 0.953 & 0.953 \\
      	& length of CI & 3.760 & 3.824 & 3.833  & 3.844 & 3.837 & 3.851  & 3.830  & 3.837    \\
\hline
      25 \% & bias & -0.015 & -0.019 & -0.018 &   -0.019   &   -0.018  &  -0.019 & -0.021 & -0.024  \\
      & MSE & 1.112 & 1.153 & 1.160 & 1.168 & 1.163 & 1.173 & 1.158 & 1.163\\
      & coverage & 0.955 & 0.966 &  0.968  & 0.970 & 0.968  & 0.970 & 0.967  &  0.969\\
      	& length of CI & 4.114 & 4.515 & 4.567 & 4.624  & 4.586 & 4.664 & 4.549  & 4.586  \\
\hline
      35 \% & bias & -0.028 & -0.029 & -0.034 &   -0.039   &   -0.036  &  -0.042 & -0.032 & -0.036  \\
      & MSE & 1.293 & 1.355 & 1.368  & 1.381 & 1.372  & 1.391 & 1.363 &1.372\\
      & coverage & 0.950 & 0.978 & 0.980 & 0.981 & 0.980 & 0.982 & 0.980 & 0.980  \\
      	& length of CI & 4.425& 5.260 & 5.362  & 5.472 &  5.398 &  5.550 & 5.327 &  5.398 \\
\hline	
\end{tabular}

\end{table}

\begin{table}[htpt]
\centering
\renewcommand{\arraystretch}{1.5}
\scriptsize
\caption{Scenario 6: Bias, MSE, coverage probability, interval length for the estimated treatment effect $\theta$, based on 1000 replications.  The 
  the sample size is 1000.}
\label{sup:scen6:N=1000}
\begin{tabular}{@{}cccccccccc@{}}
\hline missing \% & & none & point mass & unif$_{10}$ & unif$_{20}$ &   tri1$_{10}$ & tri1$_{20}$  &  tri2$_{10}$ & tri2$_{20}$    \\
\hline
      10 \% & bias & 0.003 & -0.004 & -0.006  & -0.008  & -0.007 &  -0.010 & -0.005 & -0.006 \\
      & MSE & 0.461 & 0.468 & 0.470 & 0.471 & 0.470 & 0.472 & 0.469 & 0.470\\
      & coverage & 0.956 & 0.955 & 0.955 & 0.956  & 0.955 & 0.956 & 0.955  & 0.954 \\
      	& length of CI & 2.673 & 2.723 & 2.732 &  2.742  & 2.735 &2.749  & 2.729 & 2.736    \\
\hline
      25 \% & bias & 0.023 & 0.020 & 0.015 &  0.011 &  0.014& 0.008  & 0.017 & 0.014  \\
      & MSE & 0.545 & 0.569 & 0.575  & 0.582 & 0.578  &0.587  &0.573  & 0.578\\
      & coverage & 0.956 & 0.976 &  0.977  & 0.977  &  0.977 & 0.977 &  0.976  &  0.977\\
      	& length of CI & 2.933 & 3.279 & 3.327 & 3.379  & 3.344  & 3.415 &  3.310 & 3.344  \\
\hline
      35 \% & bias & 0.019 & -0.004 & -0.012 & -0.019 & -0.014 & -0.024 & -0.009 & -0.014  \\
      & MSE & 0.604 & 0.702 & 0.720 &  0.739 & 0.726  & 0.753 & 0.713 & 0.726\\
      & coverage & 0.957 & 0.983 & 0.986 & 0.988  & 0.988  &  0.988  & 0.984  &  0.987  \\
      	& length of CI & 3.154 & 3.892 & 3.986 &  4.087 & 4.019 & 4.157  & 3.955 & 4.019 \\
\hline	
\end{tabular}

\end{table}

\begin{table}[htpt]
\centering
\renewcommand{\arraystretch}{1.5}
\scriptsize
\caption{Scenario 2: Bias, MSE, coverage probability, interval length for the estimated treatment effect $\theta$, based on 1000 replications.  The 
  the sample size is 500.}
\label{sup:scen2:N=500}
\begin{tabular}{@{}cccccccccc@{}}
\hline missing \% & & none & point mass & unif$_{10}$ & unif$_{20}$ &   tri1$_{10}$ & tri1$_{20}$  &  tri2$_{10}$ & tri2$_{20}$    \\
\hline
      10 \% & bias & 0.050 & 0.055 & 0.053 & 0.052 & 0.053 & 0.052 & 0.054 & 0.053 \\
      & MSE & 0.985 & 0.990 & 0.990 & 0.991 & 0.991 & 0.992  & 0.990  & 0.991\\
      & coverage & 0.938 & 0.935 & 0.935 & 0.935 & 0.935 & 0.935 & 0.935 & 0.935 \\
      	& length of CI & 3.720 & 3.777 & 3.786 & 3.795  & 3.789 & 3.803 & 3.783  & 3.789   \\
\hline
      25 \% & bias & 0.022 & 0.036 & 0.035 &   0.033   &   0.034  &  0.032 & 0.035 & 0.034  \\
      & MSE & 1.032 & 1.055 & 1.056 & 1.058 & 1.057 & 1.059 & 1.056 & 1.057\\
      & coverage & 0.942 & 0.954 &  0.955  & 0.957 & 0.956  & 0.958 & 0.955  &  0.957\\
      	& length of CI & 3.951 & 4.181 & 4.213 & 4.247  & 4.224 & 4.272 & 4.202  & 4.224  \\
\hline
      35 \% & bias & 0.003 & 0.035 & 0.032 &   0.029   &   0.031  &  0.027 & 0.033 & 0.031  \\
      & MSE & 1.183 & 1.205 & 1.211  & 1.218 & 1.213  & 1.222 & 1.209 &1.213\\
      & coverage & 0.944 & 0.972 & 0.976 & 0.980 & 0.976 & 0.981 & 0.975 & 0.977  \\
      	& length of CI & 4.205 & 4.810 & 4.882  & 4.962 &  4.908 &  5.020 & 4.857 &  4.908 \\
\hline	
\end{tabular}

\end{table}

\begin{table}[htpt]
\centering
\renewcommand{\arraystretch}{1.5}
\scriptsize
\caption{Scenario 2: Bias, MSE, coverage probability, interval length for the estimated treatment effect $\theta$, based on 1000 replications.  The 
  the sample size is 1000.}
\label{sup:scen2:N=1000}

\begin{tabular}{@{}cccccccccc@{}}
\hline missing \% & & none & point mass & unif$_{10}$ & unif$_{20}$ &   tri1$_{10}$ & tri1$_{20}$  &  tri2$_{10}$ & tri2$_{20}$    \\
\hline
      10 \% & bias & -0.014 & -0.012 & -0.013 & -0.014 & -0.013 & -0.015 & -0.013 & -0.013 \\
      & MSE & 0.491 & 0.494 & 0.494 & 0.495 & 0.495 & 0.496  & 0.494  & 0.495\\
      & coverage & 0.945 & 0.957 & 0.958 & 0.959 & 0.958 & 0.959 & 0.958 & 0.959 \\
      	& length of CI & 2.675 & 2.731 & 2.739 & 2.748  & 2.742 & 2.754 & 2.736  & 2.742   \\
\hline
      25 \% & bias & -0.024 & -0.004 & -0.005 &   -0.007   &   -0.006  &  -0.008 & -0.005 & -0.006  \\
      & MSE & 0.520 & 0.535 & 0.537 & 0.541 & 0.538 & 0.543  & 0.536  & 0.538\\
      & coverage & 0.947 & 0.958 &  0.961  & 0.961 & 0.961  & 0.962 & 0.960  &  0.961\\
      	& length of CI & 2.841 & 3.052 & 3.081 & 3.113  & 3.091 & 3.135 & 3.071  & 3.091  \\
\hline
      35 \% & bias & -0.011 & 0.019 & 0.016 &   0.013   &   0.015  &  0.011 & 0.017 & 0.015  \\
      & MSE & 0.550 & 0.604 & 0.611 & 0.620 & 0.614 & 0.626  & 0.609  & 0.614\\
      & coverage & 0.955 & 0.972 & 0.976 & 0.980 & 0.976 & 0.981 & 0.975 & 0.977  \\
      	& length of CI & 3.044 & 3.552 & 3.616 & 3.686  & 3.639 & 3.735 & 3.594  & 3.639  \\
\hline	
\end{tabular}

\end{table}

\pagebreak

\bibliographystyle{apa}

\bibliography{reference}